\documentclass[a4paper,11pt]{article}
\usepackage{jheppub}
\usepackage{amsmath,bm,amssymb,amsbsy,amsthm,mathrsfs,mathtools,enumitem,array,multirow}
\usepackage{graphicx,color,makecell}
\usepackage{ragged2e}
\justifying\let\raggedright\justifying

\allowdisplaybreaks[4]
\usepackage{slashed}

\newcommand{\btimes}{\rotatebox{30}{$\times$}}
\usepackage{tikz}
\usetikzlibrary{backgrounds}
\usetikzlibrary{shapes,matrix,trees}
\usetikzlibrary{arrows.meta}
\usetikzlibrary{positioning}			
\usetikzlibrary{calc,through}			
\usetikzlibrary{decorations.pathreplacing}     
\usepackage{pgffor}                                        
\usetikzlibrary{decorations.pathmorphing}	
\usetikzlibrary{decorations.markings}
\tikzset{
	vector/.style={decorate, decoration={snake,amplitude=2.5pt}, draw},
	provector/.style={decorate, decoration={snake,amplitude=2.5pt}, draw},
	antivector/.style={decorate, decoration={snake,amplitude=-2.5pt}, draw},
	fermion/.style={draw=black, postaction={decorate},
		decoration={markings,mark=at position .6 with {\arrow[draw=black]{>}}}},
           vL/.style={draw=ppurple, postaction={decorate},
		decoration={markings,mark=at position .6 with {\arrow[draw=ppurple]{>}}}},
	vLp/.style={draw=ppurple, postaction={decorate},
		decoration={markings,mark=at position .7 with {\arrow[draw=ppurple]{>}}}},
	NR/.style={draw=ggreen, postaction={decorate},
		decoration={markings,mark=at position .62 with {\arrow[draw=ggreen]{>}}}},	
	NRp/.style={draw=ggreen, postaction={decorate},
		decoration={markings,mark=at position .7 with {\arrow[draw=ggreen]{>}}}},	
	neutralino/.style={draw=black},
	fermionbar/.style={draw=black, postaction={decorate},
		decoration={markings,mark=at position .6 with {\arrow[draw=black]{<}}}},
	fermionnoarrow/.style={draw=black},
	gluon/.style={decorate, draw=black,
		decoration={coil,amplitude=4pt, segment length=5pt}},
	scalar/.style={dashed,draw=black, postaction={decorate},
		decoration={markings,mark=at position .55 with {\arrow[draw=black]{>}}}},
	scalarbar/.style={dashed,draw=black, postaction={decorate},
		decoration={markings,mark=at position .55 with {\arrow[draw=black]{<}}}},
	scalarnoarrow/.style={dashed,draw=black},
	electron/.style={draw=black, postaction={decorate},
		decoration={markings,mark=at position .55 with {\arrow[draw=black]{>}}}},
	bigvector/.style={decorate, decoration={snake,amplitude=4pt}, draw},
	photon/.style={decorate, draw=black,decoration={snake,amplitude=4pt, segment length=5pt} }
}
\definecolor{ggreen}{rgb}{0.0,0.4,0.2}
\definecolor{mmagenta}{rgb}{0.7,0.0,0.7}
\definecolor{ppurple}{rgb}{0.4,0.1,0.9}
\definecolor{oorange}{rgb}{0.8,0.4,0.3}
\definecolor{ggray}{rgb}{0.4,0.4,0.4}
\definecolor{bblue}{rgb}{0.2,0.5,1}
\definecolor{bbrown}{rgb}{0.6,0.3,0}
 
\definecolor{ccblue}{rgb}{0.0,0.4,0.8}
\usepackage[]{hyperref}
\hypersetup{  colorlinks=true,
	linkcolor=ccblue,
	urlcolor=ccblue,
	citecolor=ccblue}

\title{Neutrinoless Double Beta Decay in Multiple Isotopes for Fingerprints Identification of Operators and Models}
	
\author[a]{Shao-Long Chen}
\author[a]{and Yu-Qi Xiao}

\affiliation[a]{Key Laboratory of Quark and Lepton Physics (MoE) and Institute of Particle Physics, Central China Normal University, \\Wuhan 430079, China}
	
\emailAdd{chensl@ccnu.edu.cn}
\emailAdd{xiaoyq@mails.ccnu.edu.cn}

\abstract{Neutrinoless double beta ($0\nu\beta\beta$) decay is the most promising way to determine whether neutrinos are Majorana particles. There are many experiments based on different isotopes searching for $0\nu\beta\beta$ decay. Combining the searches of $0\nu\beta\beta$ decay in multiple isotopes provides a possible method to distinguish operators and different models.
The contributions to $0\nu\beta\beta$ decay come from standard, long-range, and short-range mechanisms. We analyze the scenario in which the standard and short-range operators exist simultaneously within the framework of low-energy effective field theory. Five specific models are considered, which can realize neutrino mass and can contribute to $0\nu\beta\beta$ decay via multiple mechanisms. A criterion to evaluate the possibilities of future experiments to discriminate operators and models is built. We find that the complementary searches for $0\nu\beta\beta$ decay in different isotopes can distinguish the cases that contain the low-energy effective operators $\mathcal{O}_{1,2,5}$ and R-parity violating supersymmetry model. For other cases and models, the experimental searches within multiple isotopes can also more effectively constrain the parameter region than with only one isotope.}
	
\begin{document}
\maketitle
\flushbottom

\section{Introduction}
Tiny neutrino masses are commonly assumed to be generated through the dim-5 Weinberg operator~\cite{Weinberg:1979sa} where the neutrinos are Majorana particles. The realizations can happen at the tree level and loop level. The tree-level framework is known as the seesaw mechanism~\cite{gell1979supergravity,glashow1980quarks,Foot:1988aq,Ma:1998dx,Ma:1998dn}, while the loop-level constructions have been systematically studied in~\cite{Ma:1998dn,Bonnet:2012kz}. However, neutrinos can also be Dirac particles. Therefore, it is crucial to find out what kind of particles the neutrinos are.

The neutrinoless double beta ($0\nu\beta\beta$) decay experiments offer a promising way of probing the nature of neutrinos. If neutrinos are Majorana particles, the $0\nu\beta\beta$ decay can be realized via the light neutrino exchange, and the inverse half-life of the isotopes is described by
\begin{align}
    T^{-1}_{1/2}=G_{0\nu}\bigg|\dfrac{\langle m_{ee}\rangle}{m_{e}}\mathcal{M}_{\nu}\bigg|^{2}\,,
\end{align}
which is proportional to the square of effective neutrino mass $\langle m_{ee}\rangle=\sum_{i}U_{ei}^{2}m_{\nu_{i}}$, with $U$ to be the PMNS matrix and $m_{\nu_{i}}$ to be the masses of three generations of light neutrinos. The $G_{0\nu}$ represents the phase space factor (PSF), and $\mathcal{M}_{\nu}$ is the nuclear matrix element (NME). Currently, the most stringent limits on the half-life of $0\nu\beta\beta$ decay are provided by the GERDA experiment $T_{1/2}(^{76}{\text{Ge}})>1.8\times10^{26}$ yrs~\cite{GERDA:2020xhi} and KamLAND-Zen experiment $T_{1/2}(^{136}{\text{Xe}})>1.08\times10^{26}$ yrs~\cite{KamLAND-Zen:2016pfg}. Many ton-scale experiments aim to search this decay process with different isotopes in the future, such as the LEGEND-1000 (based on $^{76}{\text{Ge}}$)~\cite{LEGEND:2021bnm}, CUPID-1T ($^{100}{\text{Mo}}$)~\cite{CUPID:2022wpt}, N$\nu$Dex ($^{82}{\text{Se}}$)~\cite{NnDEx-100:2023alw}, nEXO ($^{136}{\text{Xe}}$)~\cite{Pocar:2020zqz}, etc. One can see the details in the review, e.g.,~\cite{zel1981study,Doi:1985dx,Elliott:2002xe,Avignone:2007fu,Vergados:2012xy,Rodejohann:2012xd,Dolinski:2019nrj}.

Over the past twenty years, there has been growing interest in what searches of $0\nu\beta\beta$ decay in different isotopes can reveal. It has been pointed out in~\cite{Bilenky:2004um} that it could be a tool to solve some nuclear matrix elements (NMEs) problems.
A number of studies~\cite{Deppisch:2006hb,Gehman:2007qg,Fogli:2009py,Faessler:2011rv} have compared $0\nu\beta\beta$ decay in different new physics models with different isotopes. They demonstrated that certain models can be discriminated by future signals, and this conclusion will not be influenced by the large uncertainties on NMEs as the systematic effects can be canceled by using the ratio of half-life
\begin{align}
\dfrac{T_{1/2}(^{A}N)}{T_{1/2}(^{76}\mathrm{Ge})}=\frac{|\mathcal{M}({ }^{76} \mathrm{Ge})|^2 G\left({ }^{76} \mathrm{Ge}\right)}{|\mathcal{M}({ }^{A} N)|^2 G({ }^{A} N)}\,.
\end{align}
In addition, the authors in~\cite{Graf:2022lhj} have discussed the possibilities of distinguishing mechanisms of $0\nu\beta\beta$ decay also by comparing the half-life ratio in different isotopes with one low-energy effective operator assumed at a time.
More recently, several works have also focused on certain new physics models~\cite{Lisi:2015yma,Chen:2022rcv,Agostini:2022bjh,Lisi:2023amm,Ding:2024obt}. These works have investigated the abilities of experimental searches with multiple isotopes to discriminate the models.
The light neutrino and heavy neutrino exchange mechanisms have been considered in~\cite{Lisi:2015yma} with the main theories uncertainties being parameterized, and the authors have claimed that these two mechanisms are hard to distinguish. The R-parity violating Supersymmetry model has been discussed in~\cite{Agostini:2022bjh} with the uncertainties of NME values considered in the framework of a global Bayesian analysis.
In the literature~\cite{Lisi:2023amm}, the non-interfering case in the Left-Right symmetric model has been investigated with a large set of NME values.
The commonality of these models is that they can realize neutrino mass and contribute to $0\nu\beta\beta$ not only through light neutrino exchange but also through additional mechanisms.

Hence, in this work, we consider the cases that $0\nu\beta\beta$ decay is contributed by more than one low-energy effective operator. The aim of this paper is to give a systematic analysis to explore whether experimental searches in multiple isotopes ($^{76}\text{Ge}$, $^{82}\text{Se}$, $^{100}\text{Mo}$ and $^{136}\text{Xe}$) have the potential to identify operators and models when the theory contains more than one mechanism.

The contributions to $0\nu\beta\beta$ decay can be categorized as standard, long-range, and short-range interactions. Within the framework of effective field theory, the scenario where the short-range mechanism and light neutrino exchange mechanism co-occur is focused, and the interplay between these two mechanisms is investigated. 
To give intuitive results, the plots of correlation between Wilson coefficients with the cancellation effect are shown. The experimental limits on the half-life of isotopes can be converted into the survival bands or areas of the effective neutrino mass $\langle m_{ee}\rangle$ and the Wilson coefficients. The ratios corresponding to the slopes of the restriction bands and the tilt angles of the elliptical survival regions are given.  Instead of using the ratio of half-life as the previous works, the criterion to evaluate the abilities of future experiments to distinguish these operators is whether the slopes and tilt angles of the survival regions in different isotopes differ distinctively.

For the discussion on specific models, we choose the models that realize the tiny Majorana neutrino masses and contribute to $0\nu\beta\beta$ decay via multiple mechanisms simultaneously. The correlations between the parameters and the effective neutrino mass in different isotopes are investigated. The interference effects that have not always been dealt with in literature could play a key role in certain models so that the with and without interference cases are considered and shown.

The paper is arranged as follows. In Sec.~\ref{Sec2}, the mechanisms of $0\nu\beta\beta$ decay are briefly revisited. In Sec.~\ref{Sec3}, the operators in different isotopes with the short-range mechanism and light neutrino exchange mechanism occurrence are analyzed. In Sec.~\ref{Sec4}, specific neutrino mass models are discussed, and we give the correlations between the effective neutrino mass and other parameters by considering the $0\nu\beta\beta$ decay experimental limits of different isotopes. Finally, the results are summarized in Sec.~\ref{Sec5}.

\section{The mechanism of neutrinoless double beta decay}
\label{Sec2}
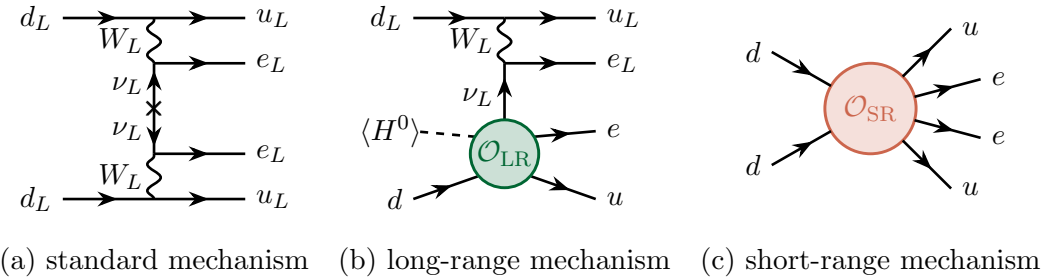
\begin{figure}[t]
\centering
\begin{tikzpicture}[line width=1pt, scale=1.5,>=Stealth]
			\path (90:0.4) coordinate (a0);
			\path (90:0.8) coordinate (a1);
			\path (-0.8,0.8) coordinate (a2);
			\path (0.8,0.8) coordinate (a3);
			\path (0.8,0.4) coordinate (a4);
			\path (270:0.4) coordinate (a5);
			\path (-0.8,-0.8) coordinate (a6);
			\path (0.8,-0.8) coordinate (a7);
			\path (0.8,-0.4) coordinate (a8);
			\path (0.8,-1.4) coordinate (a9);
			\path (0,-0.8) coordinate (a10);
			\path (0,-1.4) coordinate (a11);
			\path (0,0.4) coordinate (a12);
			\path (0,0.0) coordinate (a13);
			\path (0,-0.4) coordinate (a14);
			\draw [vector](a0)--(a1);
			\draw [fermion](a2)--(a1);
			\draw [fermion](a1)--(a3);
			\draw [fermion](a0)--(a4);
			\draw [fermion](a6)--(a10);
			\draw [fermion](a10)--(a7);
			\draw [vector](a10)--(a5);
			\draw [fermion](a5)--(a8);
		 	\draw [fermionbar](0,0.4)--(0,0);
			\draw [fermionbar](0,-0.4)--(0,0);
			\node [left] at  (90:0.6) {$W_{L}$};
			\node [right] at  (a3) {$u_{L}$};
			\node [right] at  (a4) {$e_{L}$};
			\node [right] at  (a8) {$e_{L}$};
			\node [right] at  (a7) {$u_{L}$};
			\node [left] at  (a2) {$d_{L}$};
			\node [left] at  (a6) {$d_{L}$};
			\node [left] at  (0,-0.6) {$W_L$};
			\node [left] at  (0,0.2) {$\nu_{L}$};
                                \node [left] at  (0,-0.2) {$\nu_{L}$};
			\node at (0.005,0.0) {$\pmb{\times}$};
			\node at (0,-1.35) {(a) standard mechanism};
		\end{tikzpicture}
		\begin{tikzpicture}[line width=1pt, scale=1.5,>=Stealth]
			\path (90:0.4) coordinate (a0);
			\path (90:0.8) coordinate (a1);
			\path (-0.8,0.8) coordinate (a2);
			\path (0.8,0.8) coordinate (a3);
			\path (0.8,0.4) coordinate (a4);
			\path (270:0.4) coordinate (a5);
			\path (-0.8,-1) coordinate (a6);
			\path (0.8,-1) coordinate (a7);
			\path (0.8,-0.4) coordinate (a8);
			\path (0.8,-1.4) coordinate (a9);
			\path (0,-1) coordinate (a10);
			\path (0,-1.4) coordinate (a11);
			\path (0,0.3) coordinate (a12);
			\path (0,-0.05) coordinate (a13);
			\path (0,-0.2) coordinate (a14);
			\draw [vector](a0)--(a1);
			\draw [fermion](a2)--(a1);
			\draw [fermion](a1)--(a3);
			\draw [fermion](a0)--(a4);
			\draw [fermion](-0.8,-0.8)--(-0.235,-0.6);
			\draw [fermion](0.235,-0.6)--(0.8,-0.8);
			\draw [fermion](0.25,-0.25)--(0.8,-0.2);
			\draw [scalarnoarrow](-0.25,-0.25)--(-0.8,-0.2);
			\draw [fermionbar](0,0.4)--(0,-0.1);
			\filldraw [fill=ggreen!20,draw=ggreen, line width=1.1pt] (0,-0.4) circle (0.3);
			\node [left] at  (90:0.6) {$W_{L}$};
			\node [right] at  (a3) {$u_{L}$};
			\node [right] at  (a4) {$e_{L}$};
			\node [right] at  (0.8,-0.2) {$e$};
			\node [right] at  (0.8,-0.8) {$u$};
			\node [left] at  (a2) {$d_{L}$};
			\node [left] at  (-0.8,-0.8) {$d$};
			\node at  (-1,-0.2) {$\langle H^{0}\rangle$};
			\node [left] at  (0,0.1) {$\nu_{L}$};
			\node at (0,-1.35) {(b) long-range mechanism};
			\node at (0,-0.4) {\textcolor{ggreen}{$\mathcal{O}_{\text{LR}}$}};
		\end{tikzpicture}
		\begin{tikzpicture}[line width=1pt, scale=1.5,>=Stealth]
			\filldraw [fill=oorange!20,draw=oorange,line width=1.1pt] (0,0) circle (0.4);
			\draw [fermion] (15:0.4)--(15:1);
			\draw [fermion] (-15:0.4)--(-15:1);
			\draw [fermion] (45:0.4)--(45:1);
			\draw [fermion] (-45:0.4)--(-45:1);
			\draw [fermionbar] (150:0.4)--(150:1);
			\draw [fermionbar] (210:0.4)--(210:1);
			\node [right] at  (15:1) {$e$};
			\node [right] at  (-15:1) {$e$};
			\node [right] at  (45:1) {$u$};
			\node [right] at  (-45:1) {$u$};
			\node [left] at  (150:1) {$d$};
			\node [left] at  (210:1) {$d$};
			\node at (0,-1.35) {(c) short-range mechanism};
			\node at (0,0) {\textcolor{oorange}{$\mathcal{O}_{\text{SR}}$}};
		\end{tikzpicture}
		\caption{The mechanisms of neutrinoless double beta decay.}
		\label{mechanisms}
\end{figure}

The contributions to $0\nu\beta\beta$ decay come from three mechanisms: standard mechanism, long-range mechanism, and short-range mechanism, as shown in Fig.~\ref{mechanisms}.%
~%
 The long-range and short-range correspond to the light and heavy of the exchanged particle, respectively. The contribution from the standard mechanism to the inverse half-life is proportional to the effective neutrino mass. The long-range mechanism usually refers to the cases that induce light neutrino momentum from the propagator, while the short-range mechanism contains heavy particle exchange. In this section, we briefly revisit the parameterization of the long-range and short-range mechanisms.

\paragraph{The long-range mechanism}
The long-range part of $0\nu\beta\beta$ decay has been parameterized in~\cite{Pas:1999fc} with the effective Lagrangian written as
\begin{align}
\mathcal{L}_{\text{LR}}=\frac{G_{F}}{\sqrt{2}}\left(j^{\mu}_{V-A} J_{V-A,\mu}^{\dagger}+\sum_{\alpha,\beta} \epsilon^{\beta}_{\alpha}j_{\beta}J^{\dagger}_{\alpha}\right)\,.  
\end{align}
The $j$ denotes leptonic currents which can be written as $j_{V\pm A}^{\mu}=\bar{e}\gamma^{\mu}(1\pm\gamma_{5})\nu$, $j_{S\pm P}=\bar{e}(1\pm\gamma_{5})\nu$ and $j_{T_{R/L}}^{\mu\nu}=\bar{e}\sigma^{\mu\nu}(1\pm\gamma_{5})\nu$ with $\nu=\nu_{L}+\nu_{L}^{c}$. Similarly, the hadronic currents are described as $J_{V\pm A}^{\mu}=\bar{u}\gamma^{\mu}(1\pm\gamma_{5})d$, $J_{S\pm P}=\bar{u}(1\pm\gamma_{5})d$ and $J_{T_{R/L}}^{\mu\nu}=\bar{u}\sigma^{\mu\nu}(1\pm\gamma_{5})d$. The products of leptonic and hadronic currents $j_{\beta}J_{\alpha}^{\dagger}$ are dimension-six. The Lagrangian actually could also contain the dimension-seven terms $[\overline{u}\gamma^{\mu}(1\pm\gamma_{5})d][\overline{e}\stackrel{\leftrightarrow}{\partial}_\mu(1\pm\gamma_{5})\nu^c]$, but these terms are not relevant to our later discussion. The left-handed leptonic currents induce the neutrino propagator proportional to neutrino mass, while the right-handed currents induce neutrino momentum. Attention is usually paid to the contribution of neutrino momentum, while the contribution of neutrino mass in the long-range mechanism is considered negligible compared with the standard mechanisms. To induce the current products in the long-range mechanism, heavy particles are typically introduced with masses heavier than the electroweak scale, necessitating consideration of the QCD running effects~\cite{Arbelaez:2016zlt}.

\paragraph{The short-range mechanism}
The short-range mechanism is realized by heavy particle exchange. The effective Lagrangian of the short-range mechanism is generally parameterized as the products of two hadronic currents and a leptonic current~\cite{Pas:2000vn,Deppisch:2020ztt}
\begin{align}
\mathcal{L}_{\text{SR}}
=\dfrac{G_{F}^{2}V_{ud}^{2}}{2m_{p}}
\sum\limits_{X,Y,Z}&\bigg(\epsilon_{1}^{XYZ}J_{X}J_{Y}j_{Z}
+\epsilon_{2}^{XYZ}J^{\mu\nu}_{X}J_{Y,\mu\nu}j_{Z}
+\epsilon_{3}^{XYZ}J^{\mu}_{X}J_{Y,\mu}j_{Z}\notag\\
&+\epsilon_{4}^{XY}J^{\mu}_{X}J_{Y,\mu\nu}j^{\nu}
+\epsilon_{5}^{XY}J^{\mu}_{X}J_{Y}j_{\mu}\bigg)+{\text{h.c.}}\,,\label{L_SR} 
\end{align}    
where $G_{F}$ is the Fermi constant, $m_{p}$ is the proton mass, and $X,Y,Z$ denote the chirality of the currents. The coefficients $\epsilon_{i}^{XY(Z)}$ are dimensionless Wilson coefficients at the mass scale of the introduced heavy particle. The $J$ and $j$, respectively, denote the hadronic and leptonic currents as
\begin{equation}
\begin{gathered}
J_{R/L}=\overline{u}(1\pm\gamma_{5})d\,,\quad J_{R,L}^{\mu}=\overline{u}\gamma^{\mu}(1\pm\gamma_{5})d\,,\quad J_{R/L}^{\mu\nu}=\overline{u}\sigma^{\mu\nu}(1\pm\gamma_{5})d\,,\\
j_{R/L}=\overline{e}(1\mp\gamma_{5})e^{c},\quad j^{\mu}=\overline{e}\gamma^{\mu}\gamma_{5}e^{c}\,,
\end{gathered}
\end{equation}
where the convention of the chirality of the scalar leptonic current is followed from~\cite{Deppisch:2020ztt}. One can further express the effective operators in terms of the currents as
\begin{equation}
\begin{gathered}
\mathcal{O}^{XYZ}_{1}\equiv J_{X}J_{Y}j_{Z}\,,\quad
\mathcal{O}^{XYZ}_{2}\equiv J^{\mu\nu}_{X}J_{Y,\mu\nu}j_{Z}\,,\quad
\mathcal{O}^{XYZ}_{3}\equiv J^{\mu}_{X}J_{Y,\mu}j_{Z}\,,\\
\mathcal{O}^{XY}_{4}\equiv J^{\mu}_{X}J_{Y,\mu\nu}j^{\nu}\,,\quad
\mathcal{O}^{XY}_{5}\equiv J^{\mu}_{X}J_{Y}j_{\mu}\,.
\end{gathered}
\end{equation}
which are related to the NMEs $\mathcal{M}_{i}^{XY(Z)}$. There are 24 dimension-nine operators listed in~\cite{Prezeau:2003xn,Graesser:2016bpz,Cirigliano:2018yza,Graf:2018ozy,Scholer:2023bnn}.

In a specific UV model, the contribution usually comes from multiple mechanisms instead of a single mechanism. Therefore, it is crucial to investigate various mechanisms simultaneously. In the following section, we take the standard and short-range mechanisms, for instance, to show the interplay.

\section{Interplay between standard and short-range mechanisms}\label{Sec3}
 Besides the standard mechanism, the contributions to $0\nu\beta\beta$ decay in neutrino mass models can come from other mechanisms, e.g., the short-range mechanism. We consider $0\nu\beta\beta$ decay in the framework involving the short-range mechanism and light-neutrino exchange. The following expression gives the isotope $0\nu\beta\beta$ decay inverse half-life~\cite{Deppisch:2020ztt}
\begin{align}
\left[T^{0\nu\beta\beta}_{1/2}\right]^{-1}
&=G_{11+}^{(0)}\left|\sum\limits_{i=1}^{3}\epsilon_{i}^{XYL}\mathcal{M}^{XYL}_{i}+\epsilon_{\nu}\mathcal{M}_{\nu}\right|^{2}
+G_{11+}^{(0)}\left|\sum\limits_{i=1}^{3}\epsilon_{i}^{XYR}\mathcal{M}^{XYR}_{i}\right|^{2}
+G_{66}^{(0)}\left|\sum\limits_{i=4}^{5}\epsilon_{i}^{XY}\mathcal{M}^{XY}_{i}\right|^{2}\notag\\
&+G_{16}^{(0)}\times 2{\text{Re}}\left[\left(\sum\limits_{i=1}^{3}\epsilon_{i}^{XYL}\mathcal{M}^{XYL}_{i}-\sum\limits_{i=1}^{3}\epsilon_{i}^{XYR}\mathcal{M}^{XYR}_{i}+\epsilon_{\nu}\mathcal{M}_{\nu}\right)\left(\sum\limits_{i=4}^{5}\epsilon_{i}^{XY}\mathcal{M}^{XY}_{i}\right)^{*}\right]\notag\\
&+G_{11-}^{(0)}\times2 {\text{Re}}\left[\left(\sum\limits_{i=1}^{3}\epsilon_{i}^{XYL}\mathcal{M}^{XYL}_{i}+\epsilon_{\nu}\mathcal{M}_{\nu}\right)\left(\sum\limits_{i=1}^{3}\epsilon_{i}^{XYR}\mathcal{M}^{XY}_{i}\right)^{*}\right]\,,
\label{0vbb}
\end{align}
where the dimensionless parameter $\epsilon_{\nu}$ equals $\langle m_{ee}\rangle/m_{e}$ with $\langle m_{ee}\rangle\equiv\sum_{i}U_{ei}^{2}m_{i}$ denoting the effective neutrino mass, and $\mathcal{M}^{XYZ}_{i},\mathcal{M}_{\nu}$ are the NMEs with the short-range mechanism and the light neutrino exchange, respectively. The values of the NMEs and the PSFs have been completely listed in~\cite{Deppisch:2020ztt}, and the QCD running effects have been discussed in~\cite{Gonzalez:2015ady}.
 
The current null results in experimental searches can be represented as survival regions of the parameters. If the $0\nu\beta\beta$ decay dominantly contributed by the standard mechanism with a single short-range operator, one can discover that the relations between effective neutrino mass $\langle m_{ee}\rangle$ and the coefficient $\epsilon_{i}^{XYZ}$ can be divided into two cases, linear and elliptical case. The linear and elliptical describe the shapes of survival regions. The first term contains a linear combination of $\epsilon_{i}^{XYL}$ and $\epsilon_{\nu}$, while other terms are the bilinear combinations that induce the elliptical region.

\paragraph{Linear case} For the linear case, the survival region under the standard mechanism with a single operator dominance assumption is described by 
\begin{align}
\bigg|\epsilon_{i}^{XYL}\dfrac{\mathcal{M}_{i}^{XYL}}{\mathcal{M}_{\nu}}+\dfrac{\langle m_{ee}\rangle}{m_{e}}\bigg|^{2}<\dfrac{[T_{1/2}^{\text{exp}}]^{-1}}{G_{11}^{(0)}\mathcal{M}_{\nu}^{2}}\,,\label{linear}
\end{align}
with the experimental constraint $T_{1/2}^{\text{exp.}}$. As the effective neutrino mass and the coefficients can be complex, they can be expressed by their modulus with phases $\langle m_{ee}\rangle=|\langle m_{ee}\rangle|e^{i\alpha},\epsilon_{i}^{XYL}(\mathcal{M}_{i}^{XYL}/\mathcal{M}_{\nu})=R_{i}^{XY}|\epsilon_{i}^{XYL}|e^{i\beta}$ where $\alpha,\beta\in[0,2\pi)$, and the ratio $R_{i}^{XY}=|\mathcal{M}_{i}^{XYL}/{\mathcal{M}_{\nu}}|$. Then the left-hand side in Eq.~(\ref{linear}) can be written as
\begin{align}
|\epsilon_{i}^{XYL}|^{2}(R_{i}^{XY})^{2}+|\langle m_{ee}\rangle|^{2}/m_{e}^{2}+2R_{i}^{XY}\cos(\alpha-\beta)\cdot|\epsilon_{i}^{XYL}|\cdot|\langle m_{ee}\rangle|/m_{e}\,,
\end{align}
where $\alpha-\beta$ also varies from 0 to 2$\pi$. When $\alpha-\beta$ equals $\pi$, the survival region is maximum and exactly the linear band where the slope is determined by the ratio $R_{i}^{XY}$, which varies from different isotopes. Comparing the values of the $R_{i}^{XY}$ in different isotopes can help us to determine whether the corresponding operators are easier to distinguish in experiments, i.e., if the $R_{i}^{XY}$ values are quite different in different isotopes, then the experimental searches have a much larger opportunity to distinguish the operator $\mathcal{O}^{XYL}_{i}$ and comprehensive constraint the parameter space. The conclusion holds in different methods of NME calculations, as the systematic effects have been reduced with the NME ratio taken
\begin{align}
\dfrac{R_{i}^{XY}(^{A}N_{1})}{R_{i}^{XY}(^{B}N_{2})}=\left|\dfrac{\mathcal{M}_{i}^{XYL}(^{A}N_{1})}{\mathcal{M}_{i}^{XYL}(^{B}N_{2})}\right|\cdot\left|\dfrac{\mathcal{M}_{\nu}(^{B}N_{2})}{\mathcal{M}_{\nu}(^{A}N_{1})}\right|\,.
\end{align}

\begin{figure}[bht]
\centering
\includegraphics[height=4.8cm]{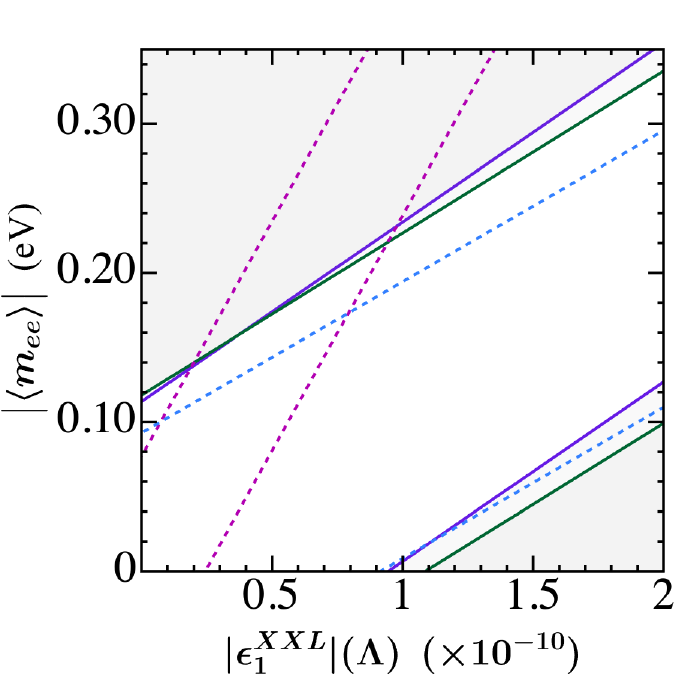}\quad
\includegraphics[height=4.8cm]{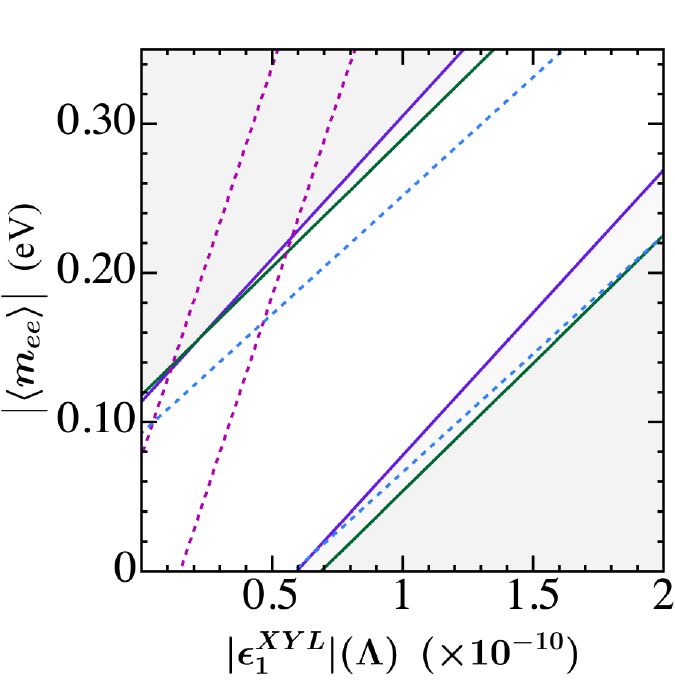}\quad
\includegraphics[height=4.8cm]{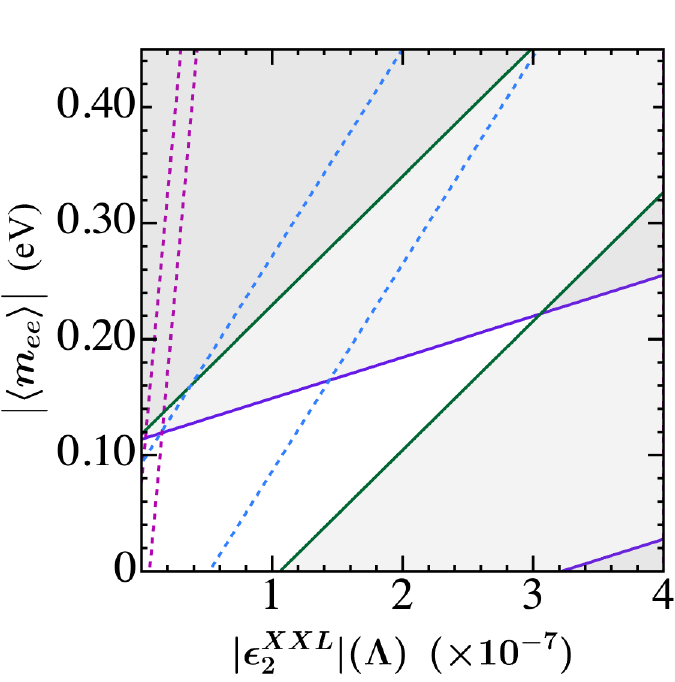}\\\vskip 0.2cm
\includegraphics[height=4.8cm]{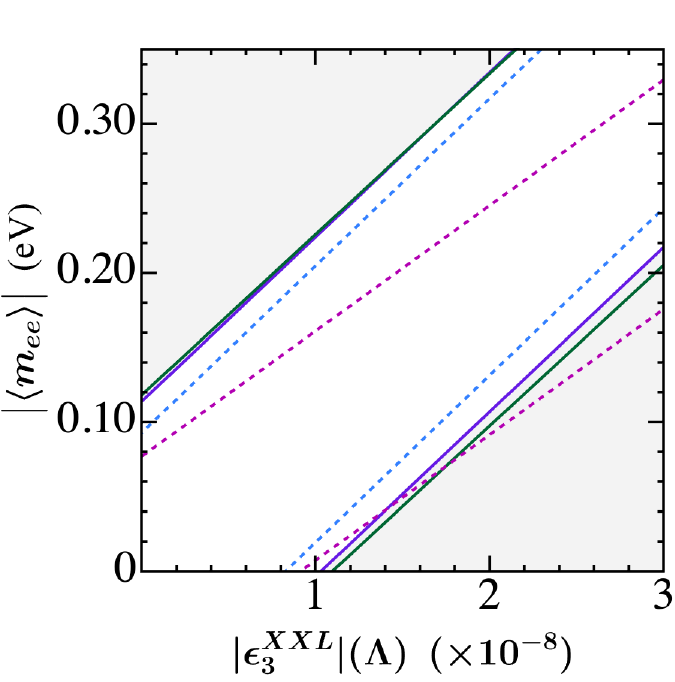}\quad
\includegraphics[height=4.8cm]{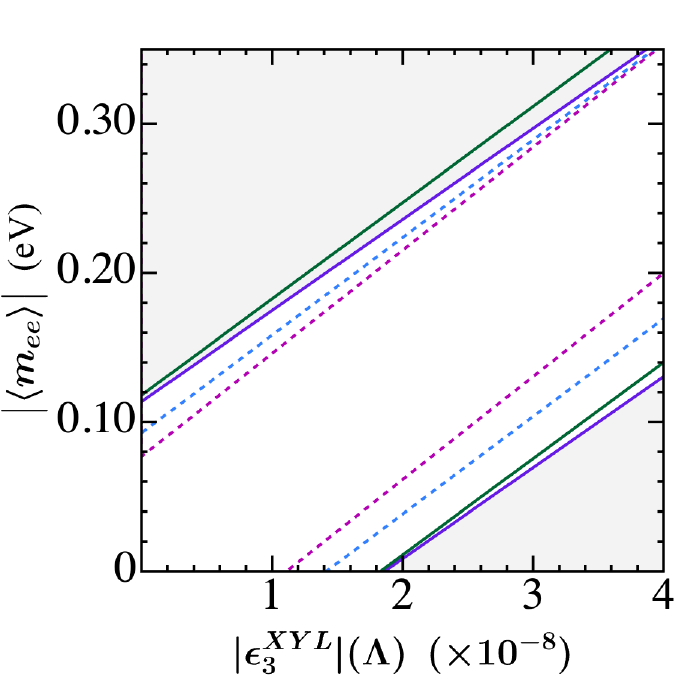}\quad
\includegraphics[height=3.8cm]{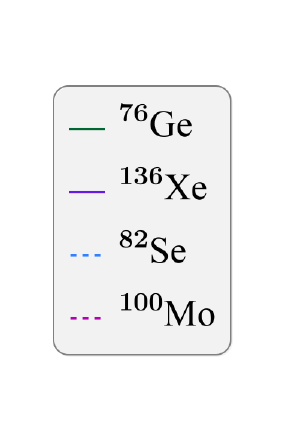}
\caption{The contours of effective neutrino mass and Wilson coefficients in isotopes $^{76}{\rm{Ge}},~^{82}{\rm{Se}},~^{100}{\rm{Mo}}$, and $^{136}{\rm{Xe}}$ with the energy scale of the coefficients $\Lambda\sim1$~TeV. The solid green and purple lines correspond to the GERDA and KamLAND-Zen experimental results $T_{1/2}({\text{GERDA}})>1.8\times 10^{26}~\text{yrs}$,~$T_{1/2}({\text{KamLAND-Zen}})>1.07\times 10^{26}~\text{yrs}$, and the gray regions are excluded. The dashed blue and magenta lines correspond to the isotopes $^{82}$Se and $^{100}$Mo with the half-life set to be $10^{26}$ yrs.}\label{linear-plot}
\end{figure}

\begin{table}[t]
\centering
\begin{tabular}{c|cccc}
\hline\hline
Ratios & $^{76}{\rm{Ge}}$ & $^{82}{\rm{Se}}$ & $^{100}{\rm{Mo}}$ & $^{136}{\rm{Xe}}$ \\
\hline
$R_{1}^{XX}$ & 806 & 746 & 2467 & 900 \\
$R_{1}^{XY}$ & 805 & 745 & 2467 & 899 \\
$R_{2}^{XX}$ & 52 & 53 & 72 & 54 \\
$R_{3}^{XX}$ & 30 & 31 & 24 & 31 \\
$R_{3}^{XY}$ & 15 & 15 & 16 & 14 \\
\hline
\hline
\end{tabular}\qquad\qquad
\begin{tabular}{c|cccc}
\hline\hline
Ratios & $^{76}{\rm{Ge}}$ & $^{82}{\rm{Se}}$ & $^{100}{\rm{Mo}}$ & $^{136}{\rm{Xe}}$ \\
\hline
$R_{1}^{XX}$ & 2126 & 1983 & 6171 & 2355 \\
$R_{1}^{XY}$ & 3359 & 3111 & 10224 & 3745 \\
$R_{2}^{XX}$ & 2.2 & 3.5 & 24.3 & 0.7 \\
$R_{3}^{XX}$ & 21 & 22 & 17 & 22 \\
$R_{3}^{XY}$ & 13 & 13 & 14 & 12 \\
\hline
\hline
\end{tabular}
\caption{The ratio values of different NMEs in isotopes $^{76}{\rm{Ge}},^{82}{\rm{Se}},^{100}{\rm{Mo}}$, and $^{136}{\rm{Xe}}$ without (left) and with (right) QCD running effect considered.}\label{linear-ratio1}
\end{table}%
\begin{figure}[b]
\centering
\includegraphics[height=5.2cm]{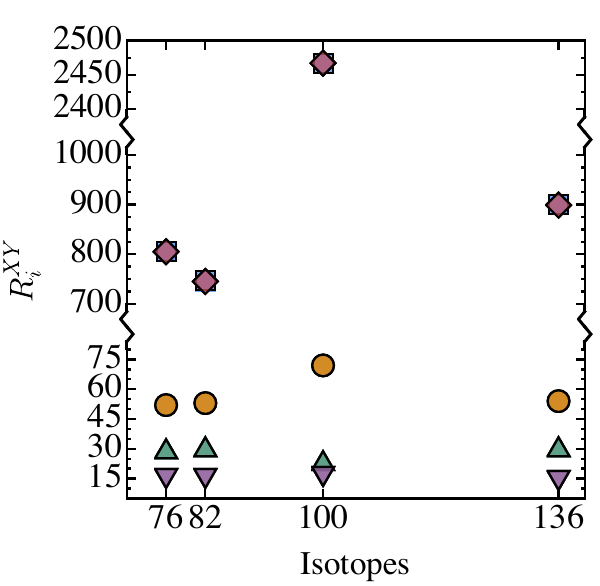}\quad
\includegraphics[height=5.2cm]{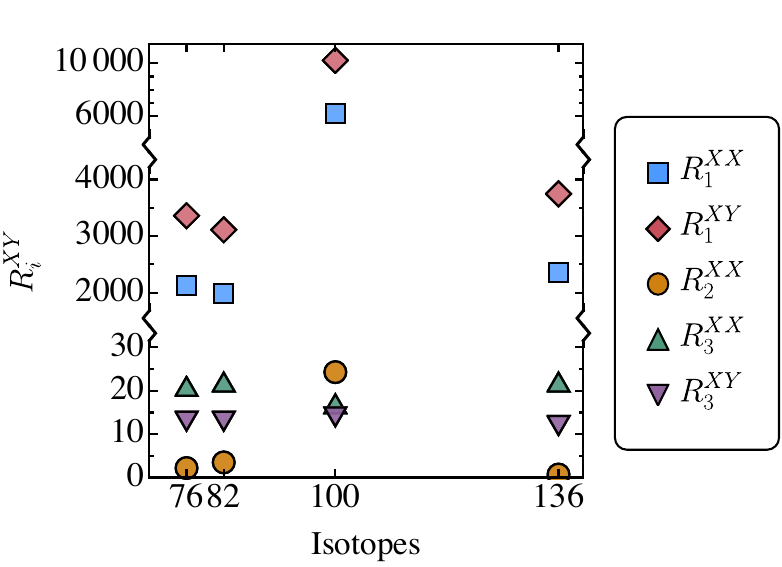}
\caption{The visualization of the ratio values of different NMEs in isotopes $^{76}{\rm{Ge}},~^{82}{\rm{Se}},~^{100}{\rm{Mo}}$, and $^{136}{\rm{Xe}}$ without (left) and with (right) QCD running effect considered.}\label{linear-ratio2}
\end{figure}
We show the ratio values for various isotopes in Table~\ref{linear-ratio1} and the visualization of these values in Fig.~\ref{linear-ratio2}. One can find that the $R_{1}^{XX}(^{100}{\rm{Mo}}), ~R_{1}^{XY}(^{100}{\rm{Mo}})$ are around three times larger than the values of other isotopes and $R_{2}^{XX}(^{100}{\rm{Mo}})$ reaches tens of times with the effects of QCD running included where the high-energy scale is around $\Lambda\sim1~\text{TeV}$. This is a fascinating thing that indicates operators $\mathcal{O}_{1,2}^{XX}$ and $\mathcal{O}_{1}^{XY}$ around TeV scale can be more likely to be distinguished by experiments. The contours of the effective neutrino mass and the Wilson coefficients are shown in Fig.~\ref{linear-plot}.
In the three figures in the upper row, the slopes of the bands corresponding to different isotopes have distinct differences. The combination of experimental searches in multiple isotopes gives a more comprehensive examination of the parameter spaces for $\epsilon_{1,2}^{XXL},\epsilon_{1}^{XYL}$ cases compared to $\epsilon_{3}^{XXL},\epsilon_{3}^{XYL}$ cases. The gray regions have been excluded by the experiments GERDA and KamLAND-Zen. If there is no signal in the $^{100}$Mo (magenta dashed line) and $^{82}$Se (blue dashed line) experiments with the sensitivity of the half-life at $10^{26}$ yrs order, the parameter survival region is reduced to the overlap area of the bands. Furthermore, a scenario in which no signals are observed in $^{76}$Ge, $^{82}$Se, or $^{136}$Xe experiments, while a signal is detected in $^{100}$Mo experiments, can potentially reveal the underlying contribution from $\mathcal{O}_{1}^{XXL}$ and $\mathcal{O}_{1}^{XYL}$. 

\begin{figure}[t]
\centering
\includegraphics[height=4.5cm]{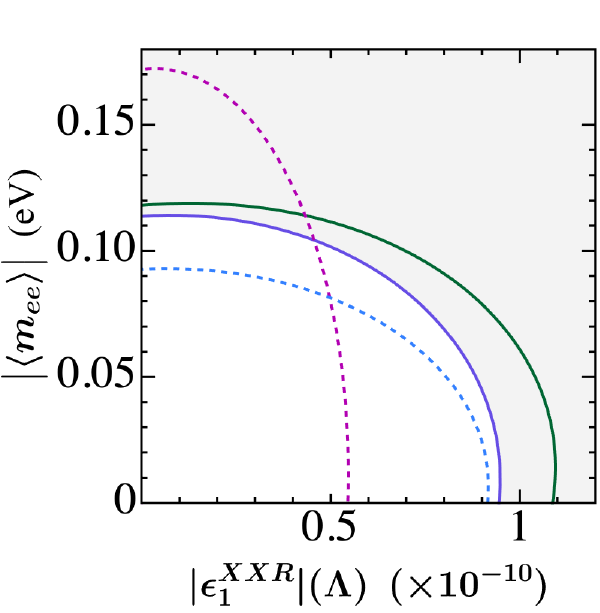}\quad
\includegraphics[height=4.5cm]{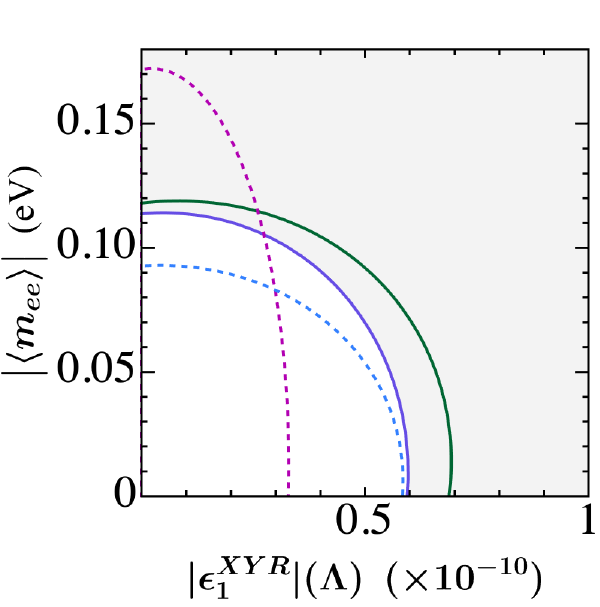}\quad
\includegraphics[height=4.5cm]{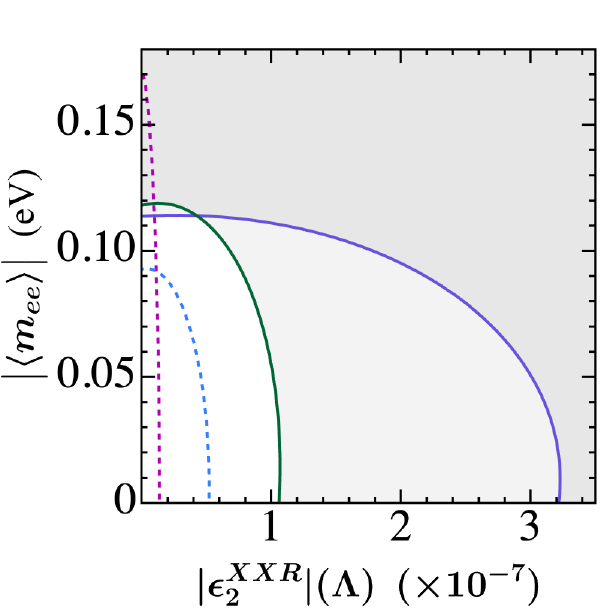}\\\vskip 0.1cm
\includegraphics[height=4.5cm]{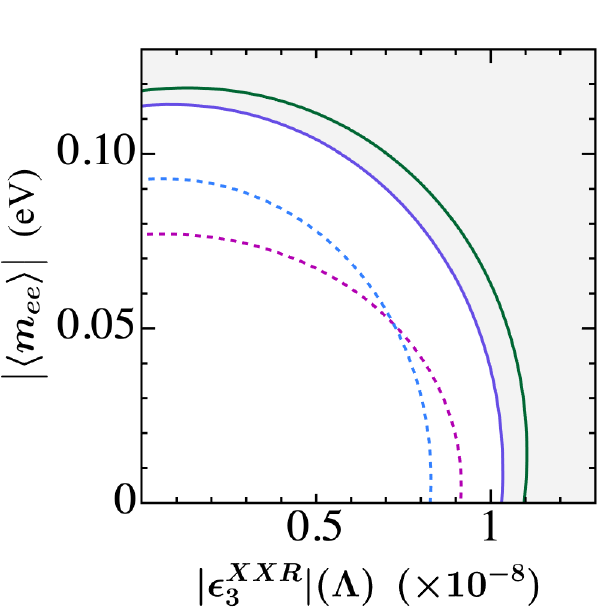}\quad
\includegraphics[height=4.5cm]{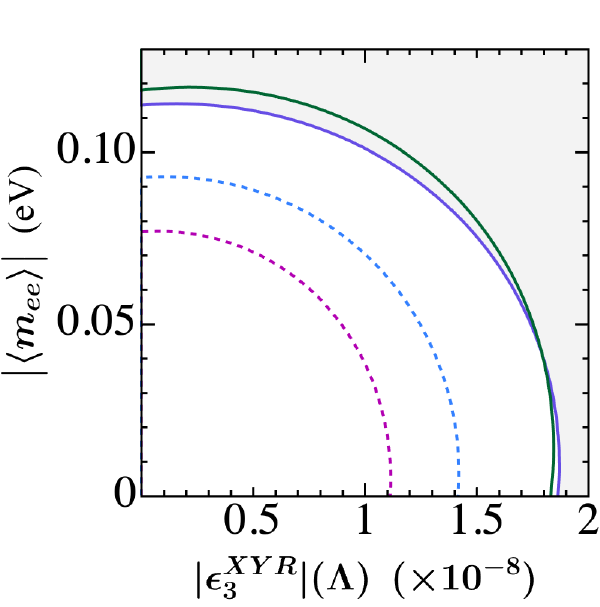}\quad
\includegraphics[height=4.5cm]{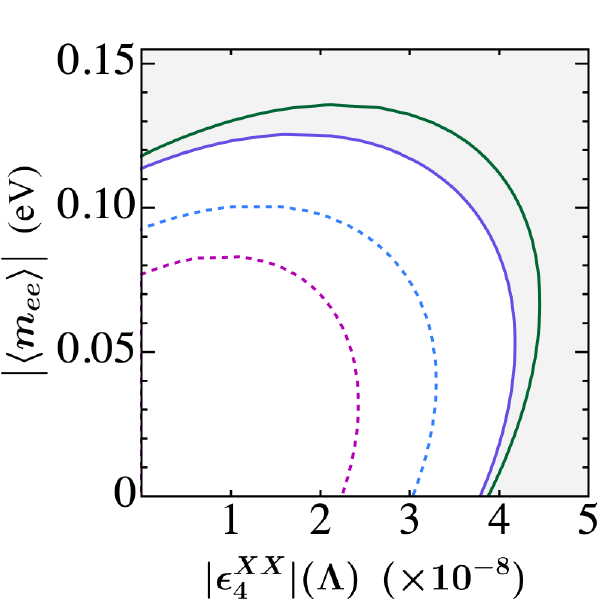}\\\vskip 0.1cm
\includegraphics[height=4.5cm]{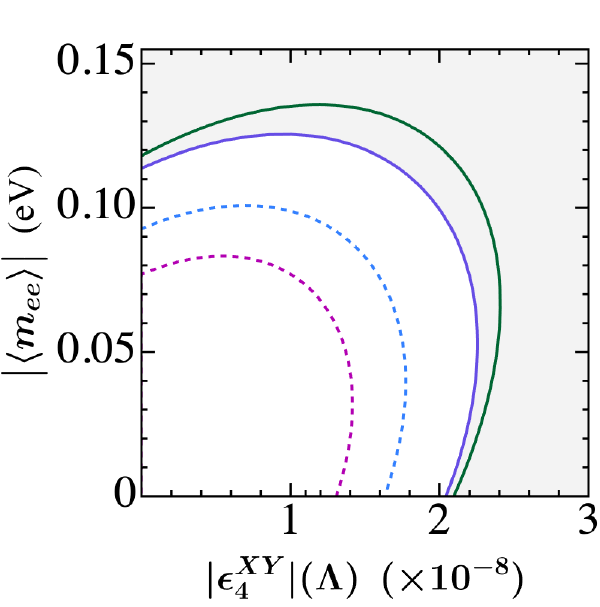}\quad
\includegraphics[height=4.5cm]{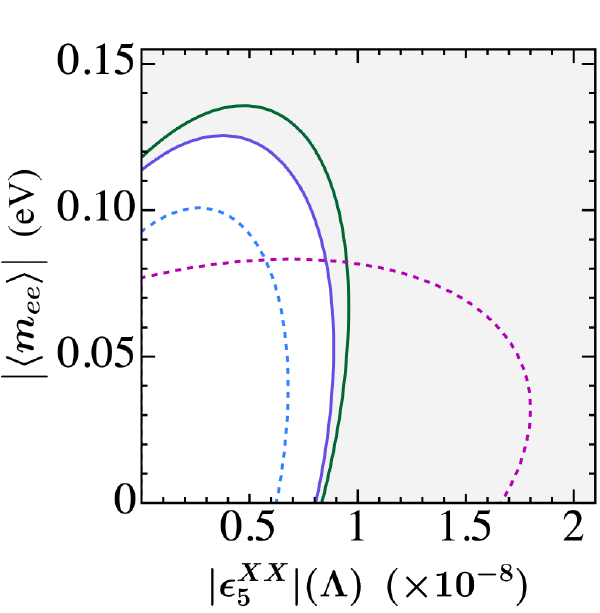}\quad
\includegraphics[height=4.5cm]{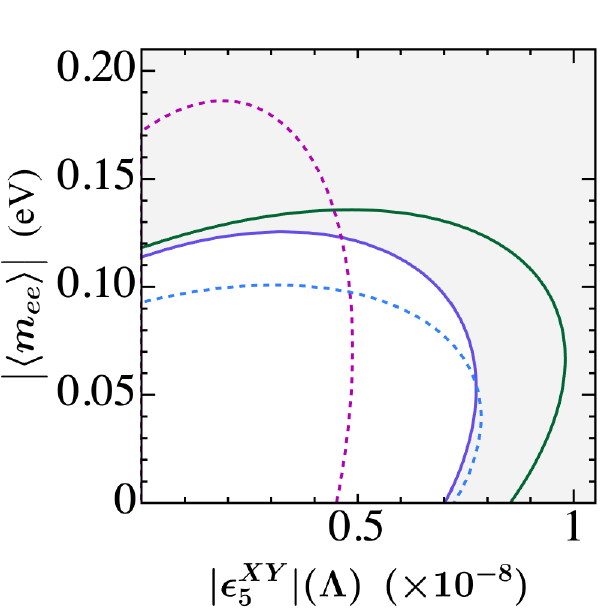}
\caption{The contours of effective neutrino mass and Wilson coefficients in isotopes $^{76}{\rm{Ge}},^{82}{\rm{Se}},^{100}{\rm{Mo}}$, and $^{136}{\rm{Xe}}$ with the energy scale of the coefficients $\Lambda\sim 1~\text{TeV}$. For the details, see the text.}\label{elliptical-plot}
\end{figure}

\paragraph{Elliptical case} For the elliptical case, one can describe the survival area via the equation of an ellipse
\begin{align}
\sqrt{(x-p)^{2}+(y-q)^{2}}+\sqrt{(x+p)^{2}+(y+q)^{2}}=2a\,,
\end{align}
with the foci $(\pm p,\pm q)$ and width $2a$. The equation can be expanded as
\begin{align}
(a^{2}-p^{2})x^{2}-2pq\cdot xy+(a^{2}-q^{2})y^{2}=a^{2}(a^{2}-p^{2}-q^{2})\,.
\end{align}
For example, the survival area for operator $\mathcal{O}_{5}^{RR}$ can be written as
\begin{align}
G_{11+}^{(0)}\bigg|\dfrac{\langle m_{ee}\rangle}{m_{e}}\mathcal{M}_{\nu}\bigg|^{2}+G_{66}^{(0)}\big|\epsilon_{5}^{RR}\mathcal{M}_{5}^{RR}\big|^{2}+2\cos{\delta}\cdot G_{16}^{(0)}\bigg|\dfrac{\langle m_{ee}\rangle}{m_{e}}\mathcal{M}_{\nu}\bigg|\cdot\big|\epsilon_{5}^{RR}\mathcal{M}_{5}^{RR}\big|<[T_{1/2}^{\text{exp}}]^{-1}\,,
\end{align}
with the phase difference $\delta$ between $\langle m_{ee}\rangle$ and $\epsilon_{5}^{RR}$ varies from 0 to $2\pi$. Then, the foci and width can be determined through
\begin{align}
\dfrac{G_{11+}^{(0)}|\mathcal{M}_{\nu}|^{2}}{a^{2}-p^{2}}
=\dfrac{G_{66}^{(0)}|\mathcal{M}_{5}^{RR}|^{2}}{a^{2}-q^{2}}
=\dfrac{G_{16}^{(0)}\cos{\delta}|\mathcal{M}_{\nu}\mathcal{M}_{5}^{RR}|}{-pq}
=\dfrac{[T_{1/2}^{\text{exp}}]^{-1}}{a^{2}(a^{2}-p^{2}-q^{2})}\,,
\end{align}
where $\delta=0$ or $\pi$ in numeraical calculation. The tilt angle $\theta$ of the survival elliptical region can be determined by $\tan{\theta}=|q/p|$, and it could be expressed as $\tan\theta=|\mathcal{R}_{5}/2+\sqrt{(\mathcal{R}_{5}/2)^{2}+1}|\simeq|1 / R_5|$ with
\begin{align}
\mathcal{R}_{5}=\dfrac{G_{11+}^{(0)}}{G_{16}^{(0)}}\left|\dfrac{\mathcal{M}_{\nu}}{\mathcal{M}_{5}^{RR}}\right|
-\dfrac{G_{66}^{(0)}}{G_{16}^{(0)}}\left|\dfrac{\mathcal{M}_{5}^{RR}}{\mathcal{M}_{\nu}}\right|\,.
\end{align}
When comparing the tilt angles in different isotopes, $\tan{\theta_{1}}/\tan{\theta_{2}}=|\mathcal{R}_{5}(^{B}N_{2})/\mathcal{R}_{5}(^{A}N_{1})|$, the ratio of the NMEs can effectively reduce the systematic uncertainty. The comparison of tilt angle provides a reliable method that enables us to determine the ability to distinguish different operators in multiple isotopes. The values of the tilt angle in different scenarios with and without QCD running effects are listed in Table~\ref{elliptical-list},
\begin{table}[t]
\centering
\begin{tabular}{c|p{0.8cm}<{\centering}p{0.8cm}<{\centering}p{0.8cm}<{\centering}p{0.8cm}<{\centering}}
\hline\hline
$\tan{\theta}$&\multirow{2}*{$^{76}{\rm{Ge}}$}&\multirow{2}*{$^{82}{\rm{Se}}$}&\multirow{2}*{$^{100}{\rm{Mo}}$}&\multirow{2}*{$^{136}{\rm{Xe}}$}\\
$(\times10^{-3})$&&&\\
\hline
$\mathcal{O}_{1}^{XXR}$&0.15&0.094&0.027&0.091\\
$\mathcal{O}_{1}^{XYR}$&0.15&0.094&0.027&0.091\\
\hline
$\mathcal{O}_{2}^{XXR}$&2.3&1.3&0.92&1.5\\
\hline
$\mathcal{O}_{3}^{XXR}$&3.9&2.2&2.8&2.7\\
$\mathcal{O}_{3}^{XYR}$&7.9&4.6&4.1&5.8\\
\hline
$\mathcal{O}_{4}^{XX}$&28&22&21&24\\
$\mathcal{O}_{4}^{XY}$&28&22&21&24\\
\hline
$\mathcal{O}_{5}^{XX}$&115&82&111&98\\
$\mathcal{O}_{5}^{XY}$&76&66&21&56\\
\hline\hline
\end{tabular}\quad\quad
\begin{tabular}{c|p{0.8cm}<{\centering}p{0.8cm}<{\centering}p{1cm}<{\centering}p{1cm}<{\centering}}
\hline\hline
$\tan{\theta}$&\multirow{2}*{$^{76}{\rm{Ge}}$}&\multirow{2}*{$^{82}{\rm{Se}}$}&\multirow{2}*{$^{100}{\rm{Mo}}$}&\multirow{2}*{$^{136}{\rm{Xe}}$}\\
$(\times10^{-3})$&&&\\
\hline
$\mathcal{O}_{1}^{XXR}$&0.056&0.035&0.011&0.035\\
$\mathcal{O}_{1}^{XYR}$&0.035&0.022&0.0065&0.022\\
\hline
$\mathcal{O}_{2}^{XXR}$&69&22&2.7&9200\\
\hline
$\mathcal{O}_{3}^{XXR}$&5.6&3.2&4.0&3.8\\
$\mathcal{O}_{3}^{XYR}$&9.5&5.5&4.9&6.9\\
\hline
$\mathcal{O}_{4}^{XX}$&82&65&57&71\\
$\mathcal{O}_{4}^{XY}$&45&35&33&62\\
\hline
$\mathcal{O}_{5}^{XX}$&18&13&42&15\\
$\mathcal{O}_{5}^{XY}$&18&16&5.1&13\\
\hline\hline
\end{tabular}
\caption{The values of the tilt angles in isotopes $^{76}{\rm{Ge}},^{82}{\rm{Se}},^{100}{\rm{Mo}}$, and $^{136}{\rm{Xe}}$ without (left) and with (right) QCD running effect considered.}\label{elliptical-list}
\end{table}
and the contour plots of effective neutrino mass and the Wilson coefficients are shown in Fig.~\ref{elliptical-plot}.

The solid green and purple lines are set by the GERDA and KamLAND-Zen results. The gray regions are excluded, and the dashed blue lines correspond to the isotopes $^{82}$Se with the half-life of $10^{26}$ yrs. In addition, the magenta lines correspond to $^{100}$Mo with the half-life to be $10^{26}$ yrs in $\epsilon_{4(3)}^{XX(R)},\epsilon_{4(3)}^{XY(R)},\epsilon_{5}^{XX}$ scenarios, while the others are with $5\times10^{26}$ yrs. We can get a similar conclusion that the searches in multiple isotopes examine the parameter regions more comprehensively than in single isotopes. Moreover, if only future $^{100}$Mo experiments have signals, it could reveal the contributions from $\mathcal{O}_{1}^{XXR},\mathcal{O}_{1}^{XYR}$, and $\mathcal{O}_{5}^{XY}$. Conversely, if only future $^{100}$Mo experiments have no signals, it may expose the $\mathcal{O}_{5}^{XX}$ contribution.

Therefore, the combined experimental searches in different isotopes have a large possibility to examine the parameter spaces and to distinguish the contributions from $\mathcal{O}_{1,2,5}$ in the short-range mechanism. In the following section, some specific neutrino mass models are concentrated on, and the correlations between the parameters and effective neutrino mass in multiple isotope searches are shown. 

\section{The Models}\label{Sec4}
In this section, various models that can generate the tiny Majorana neutrino mass and contribute to $0\nu\beta\beta$ decay simultaneously are studied in detail. For the tree-level neutrino mass models, we consider the Type-I seesaw~\cite{
Minkowski:1977sc,Mohapatra:1979ia,Gell-Mann:1979vob,Glashow:1979nm}, Type-II seesaw~\cite{Konetschny:1977bn,Cheng:1980qt,Lazarides:1980nt,Schechter:1980gr,Mohapatra:1980yp}, and the left-right symmetric model~\cite{Mohapatra:1974gc,Senjanovic:1975rk,Senjanovic:1978ev,Mohapatra:1980yp}, which contains the Type-I seesaw dominance and Type-II seesaw dominance scenarios. Two one-loop level neutrino mass models are focused, one contains leptoquarks to be scalar internal particles~\cite{Buchmuller:1986zs,Babu:1997tx,Hewett:1997ba} and the other is the R-parity violating supersymmetry model~\cite{Weinberg:1981wj,Sakai:1981pk}. The correlations between the effective neutrino mass and the parameters in these models are investigated to determine whether the combination of $0\nu\beta\beta$ decay experiments in different isotopes can distinguish these models or examine the parameter space more comprehensively.

\subsection{Seesaw Models}
\begin{figure}[t]
\centering
\includegraphics[height=4.5cm]{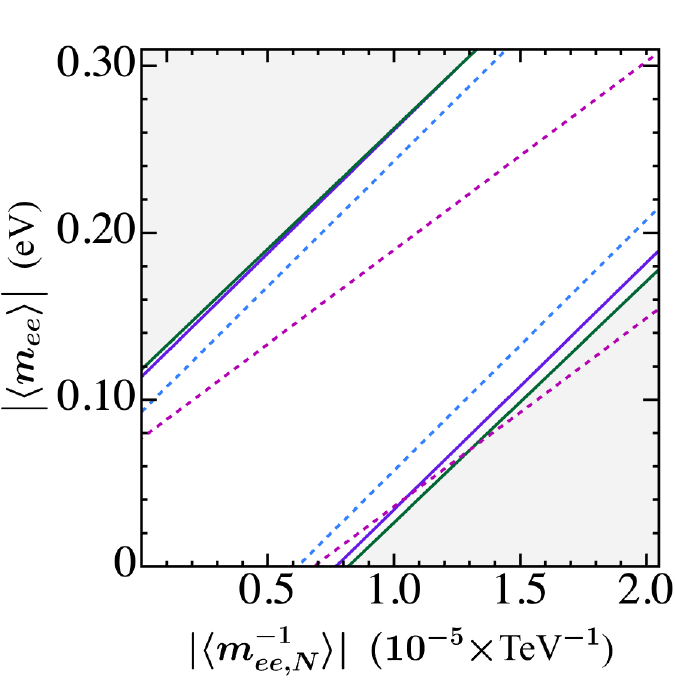}\quad
\includegraphics[height=4.5cm]{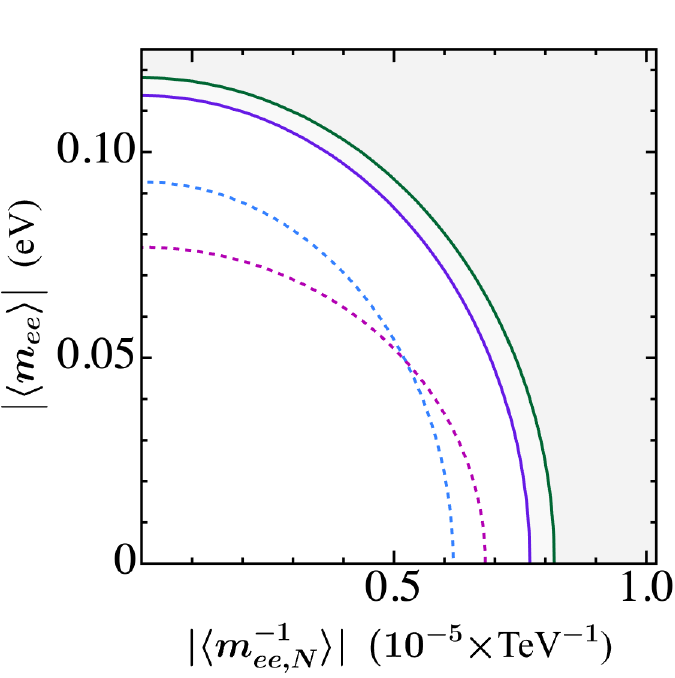}\quad
\includegraphics[height=4.5cm]{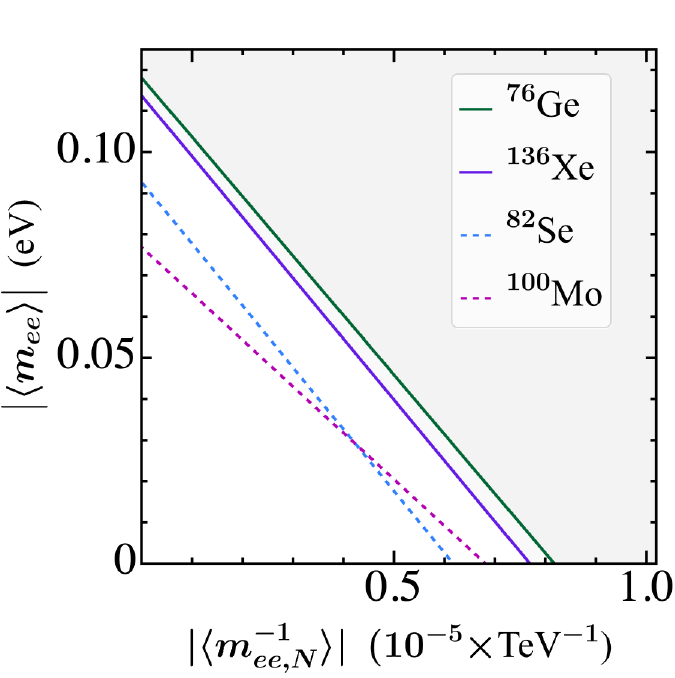}
\caption{The correlation between the effective light neutrino mass and the effective heavy neutrino mass. The phase difference in these three figures is taken to be	$\pi$ (left), $\pi/2$ (middle), and $0$ (right). The solid green and purple lines correspond to the GERDA and KamLAND-Zen experimental results, and the gray region is excluded. The dashed blue and magenta lines correspond to the isotopes $^{82}$Se and $^{100}$Mo with the half-life set to be $10^{26}$ yrs.}\label{seesawType-I}
\end{figure}

\paragraph{Type-I seesaw}
In the type-I seesaw scenario, the right-handed neutrinos (RHNs) are introduced to generate tiny neutrino mass. It has been claimed that the contributions of RHNs to $0\nu\beta\beta$ decay are important and the phenomenology could be rich with different masses of RHNs~\cite{Halprin:1983ez,Leung:1984vy,Blennow:2010th,Girardi:2013zra,Liu:2017ago,Ge:2017erv,Abada:2018qok,Bolton:2019pcu,Asaka:2020wfo,Fang:2021jfv}. Here the RHNs are considered to be much heavier than 100 MeV. The heavy neutrinos are mixed with the light active neutrinos
\begin{align}
    \nu_{\alpha L}=U_{\alpha i}\nu_{iL}+V_{\alpha j}N_{jR}^{c}\,,
\end{align}
which leads to the short-range contribution of heavy neutrino exchange 
\begin{align}
    T^{-1}_{1/2}=G_{11+}^{(0)}\bigg|\dfrac{\langle m_{ee}\rangle}{m_{e}}\mathcal{M}_{\nu}+\epsilon_{N}\mathcal{M}_{N}\bigg|^{2}\,,
\end{align}
where $\epsilon_{N}=\sum_{i}V_{ei}^{2}m_{p}/m_{N_{i}}\equiv \langle{m_{ee,N}^{-1}}\rangle m_{p}\,$. The expression of the inverse half-life is expanded as
\begin{align}
 T^{-1}_{1/2}=G_{11+}^{(0)}\bigg(\dfrac{|\langle m_{ee}\rangle|^{2}}{m_{e}^{2}}\mathcal{M}_{\nu}^{2}
 +|\epsilon_{N}|^{2}\mathcal{M}_{N}^{2}
 +2\dfrac{|\langle m_{ee}\rangle|}{m_{e}}|\epsilon_{N}|\mathcal{M}_{\nu}\mathcal{M}_{N}\cos\theta\bigg)\,,
\end{align}
where $\langle m_{ee}\rangle,~\epsilon_{N}$ have been parameterized as $\langle m_{ee}\rangle=|\langle m_{ee}\rangle|e^{i\theta_{\nu}},~\epsilon_{N}=|\epsilon_{N}|e^{i\theta_{N}}$, and the phase difference $\theta=\theta_{\nu}-\theta_{N}$ varies from 0 to $2\pi$. Fig.~\ref{seesawType-I} shows the correlation between the effective light neutrino mass and the effective heavy neutrino mass.
The left diagram shows the $\theta=\pi$ case where the cancellation appears, and the middle one shows the $\theta=\pi/2$ case without interference between the two contributions, while the right one is the $\theta=0$ case with constructive interference. It can be observed that the slope corresponding to $^{100}$Mo exhibits differences in comparison to the other isotopes. Combining experiments in multiple isotopes could restrict the parameter space more strictly.

\begin{figure}[t]
\centering
\includegraphics[height=5.2cm]{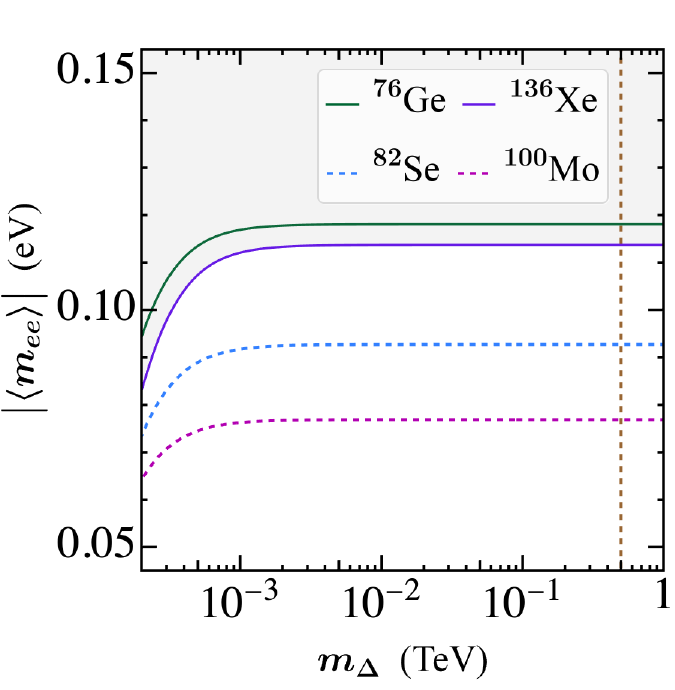}
\caption{The relation between the effective light neutrino mass and the triplet Higgs mass. The solid green and purple lines correspond to the GERDA and KamLAND-Zen experimental results, and the region above these lines is excluded. The dashed blue and magenta lines correspond to the isotopes $^{82}$Se and $^{100}$Mo with the half-life set to be $10^{26}$ yrs.}\label{seesawType-II}
\end{figure}

\paragraph{Type-II seesaw} In the type-II seesaw scenario, the Standard Model is extended by introducing a Higgs triplet $\Delta\sim(1,3,1)$. The Lagrangian containing the Higgs triplet reads
\begin{align}
\mathcal{L}&\supset{\rm{Tr}}[(D_{\mu}\Delta)^{\dagger}(D^{\mu}\Delta)]
+Y_{\Delta}^{ij}\overline{L_{L}^{ci}}i\sigma_{2}\Delta L_{L}^{j}
-m_{\Delta}^{2}{\text{Tr}}(\Delta^{\dagger}\Delta)
-\mu_{\Delta\Phi}\Phi^{T}\Delta^{*}i\sigma_{2}\Phi+{\text{h.c.}}\,,
\end{align}
with the adjoint representation 
\begin{align}
\Delta=\begin{pmatrix}\Delta^{+}/\sqrt{2}&\Delta^{++}\\\Delta^{0}&-\Delta^{+}/\sqrt{2}\end{pmatrix}\,,
\end{align}
 and the vacuum expectation values (vevs) are $\langle\phi^{0}\rangle=v/\sqrt{2}$, $\langle\Delta^{0}\rangle=v_{\Delta}/\sqrt{2}$, where $v=246~\text{GeV}$. The neutrino mass matrix is derived as $m_{\nu}^{ij}=2\mu_{\Delta\Phi}Y_{\Delta}^{ij}v^{2}/m_{\Delta}^{2}$.
The double-charged scalar contribution to neutrinoless double beta decay was first studied in literature~\cite{Mohapatra:1981pm}. The inverse half-life of the isotopes then becomes
\begin{align}
    T^{-1}_{1/2}=G_{11}^{(0)}\bigg|\dfrac{\langle m_{ee}\rangle}{m_{e}}\mathcal{M}_{\nu}+\epsilon_{\Delta}\mathcal{M}_{\Delta}\bigg|^{2}\,,
\end{align}
with the coefficient to be $\epsilon_{\Delta}=m_{\nu}^{ee}m_{p}/m_{\Delta}^{2}=\langle m_{ee}\rangle m_{p}/m_{\Delta}^{2}$ so that there is no interference difference between $\langle m_{ee}\rangle$ and $\epsilon_{\Delta}$. The relation between the effective neutrino mass $|\langle m_{ee}\rangle|$ and the mass of the Higgs triplet $m_{\Delta}$ is shown in Fig.~\ref{seesawType-II}.
The contribution from the doubly charged Higgs particle becomes significant when $m_{\Delta}<1~\text{GeV}$. However, as the ATLAS and CMS searches have put a strong limit~\cite{CMS:2017pet,CMS:2017fhs,ATLAS:2017xqs,ATLAS:2021jol}, the doubly charged Higgs mass has been determined to be heavier than 500 GeV which corresponds to the dashed brown line, and the contribution is so suppressed that one can neglect it~\cite{Schechter:1981cv,Wolfenstein:1982bf}. Therefore, it isn't easy to distinguish the Type-II seesaw model or reduce the survival regions by combined searches in multiple isotopes.

\subsection{Left-Right Symmetric Model}
In the manifest left-right symmetric model~\cite{Mohapatra:1974gc,Senjanovic:1975rk,Senjanovic:1978ev,Mohapatra:1980yp}, the gauge group is given by $SU(3)_{C}\times SU(2)_{L}\times SU(2)_{R}\times U(1)_{B-L}$ with $Q=I_{3,L}+I_{3,R}+(B-L)/2$ and couplings to be equal in left sector and right sector $g_{L}=g_{R}\equiv g$. The right-handed neutrino $N_{R}$ is introduced, and it forms an $SU(2)_{R}$ doublet with right-handed charged lepton. The doublets of the fermions are 
\begin{align}
\begin{gathered}
L_{L}=\begin{pmatrix}\nu_{L}\\ E_{L}\end{pmatrix}\sim(1,2,1,-1)\,,\quad 
L_{R}=\begin{pmatrix}N_{R}\\ E_{R}\end{pmatrix}\sim(1,1,2,-1)\,,\quad\\
Q_{L}=\begin{pmatrix}U_{L}\\ D_{L}\end{pmatrix}\sim(3,2,1,1/3)\,,\quad 
Q_{R}=\begin{pmatrix}U_{R}\\ D_{R}\end{pmatrix}\sim(3,1,2,1/3)\,.
\end{gathered}
\end{align}
In the scalar sector, the model contains a Higgs doublet $\Phi\sim(1,2,2,0)$ and two Higgs triplets $\Delta_{L}\sim(1,3,1,2)~,\Delta_{R}\sim(1,1,3,2)$
\begin{align}
\Phi=\begin{pmatrix}\phi_{1}^{0}&\phi_{2}^{+}\\\phi_{1}^{-}&\phi_{2}^{0}\end{pmatrix}\,,\quad
\Delta_{L}=\begin{pmatrix}\Delta_{L}^{+}/\sqrt{2}&\Delta_{L}^{++}\\\Delta_{L}^{0}&-\Delta_{L}^{+}/\sqrt{2}\end{pmatrix}\,,\quad
\Delta_{R}=\begin{pmatrix}\Delta_{R}^{+}/\sqrt{2}&\Delta_{R}^{++}\\\Delta_{R}^{0}&-\Delta_{R}^{+}/\sqrt{2}\end{pmatrix}\,,\quad
\end{align}
with the vevs of these scalar fields to be $\langle\phi_{1,2}^{0}\rangle=\kappa_{1,2}/\sqrt{2}$ and $\langle\Delta_{L,R}^{0}\rangle=v_{L,R}/\sqrt{2}$ where $v_{L}^{2}\ll \kappa_{1}^{2}+\kappa_{2}^{2}=v^{2}\ll v_{R}^{2}$ and $v=246~{\text{GeV}}$. The Yukawa interactions between leptons and scalars are given by 
\begin{align}
\mathcal{L}_{Y}
&=Y_{\Delta_{L}}^{ij}\overline{L_{L}^{ci}}i\sigma_{2}\Delta_{L}L_{L}^{j}
+Y_{\Delta_{R}}^{ij}\overline{L_{R}^{ci}}i\sigma_{2}\Delta_{R}L_{R}^{j}+Y_{\Phi \ell}^{ij}\overline{L_{L}^{i}}\Phi L_{R}^{j}
-\tilde{Y}_{\Phi \ell}^{ij}\overline{L_{L}^{i}}i\sigma_{2}\Phi^{*}i\sigma_{2}L_{R}^{j}+\text{h.c.}\,,
\end{align}
where the upper index $i,j=1,2,3$ denotes the generation of leptons. In this model, the tiny neutrino mass is generated by a combination of Type-I and Type-II seesaw mechanisms, and the mass matrix in the flavor basis $(\nu_{L},N_{R}^{c})$ is
\begin{align}
M_{\nu N}=
\begin{pmatrix}
M_{\Delta_{L}}&M_{\Phi\ell}\\M_{\Phi\ell}^{T}&M_{\Delta_{R}}
\end{pmatrix}
=\begin{pmatrix}
\sqrt{2}Y_{\Delta_{L}}^{\dagger}v_{L}&(Y_{\Phi\ell}\kappa_{1}+\tilde{Y}_{\Phi\ell}\kappa_{2})/\sqrt{2}\\
(Y_{\Phi\ell}^{T}\kappa_{1}+\tilde{Y}_{\Phi\ell}^{T}\kappa_{2})/\sqrt{2}&\sqrt{2}Y_{\Delta_{R}}^{\dagger}v_{R}
\end{pmatrix}\,,
\end{align}
which can be diagonalized by a unitary matrix $\tilde{U}^{\dagger}M_{\nu N}\tilde{U}^{*}=\hat{M}_{\nu N}={\text{Diag}}\{m_{1},m_{2},...,m_{6}\}$ with the
\begin{align}
\tilde{U}=\begin{pmatrix}U&V\\T&S
\end{pmatrix}\,.
\end{align}
The $0\nu\beta\beta$ decay in this model has been discussed in many literature, e.g.~\cite{Doi:1981mi, Tello:2010am, BhupalDev:2014qbx, deVries:2022nyh,Banerjee:2023aro}, where the scenarios are divided into type-I dominance and type-II dominance. 

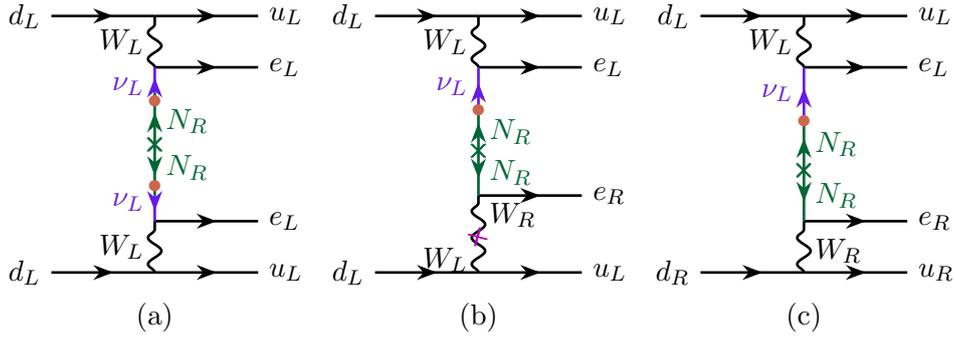
\begin{figure}[t]
\centering
\begin{tikzpicture}[line width=1pt, scale=1.7,>=Stealth]
			\path (90:0.6) coordinate (a0);
			\path (90:1) coordinate (a1);
			\path (-0.8,1) coordinate (a2);
			\path (0.8,1) coordinate (a3);
			\path (0.8,0.6) coordinate (a4);
			\path (270:0.6) coordinate (a5);
			\path (-0.8,-1) coordinate (a6);
			\path (0.8,-1) coordinate (a7);
			\path (0.8,-0.6) coordinate (a8);
			\path (0.8,-1.4) coordinate (a9);
			\path (0,-1) coordinate (a10);
			\path (0,-1.4) coordinate (a11);
			\path (0,0.4) coordinate (a12);
			\path (0,0.0) coordinate (a13);
			\path (0,-0.4) coordinate (a14);
			\draw [vector](a0)--(a1);
			\draw [fermion](a2)--(a1);
			\draw [fermion](a1)--(a3);
			\draw [fermion](a0)--(a4);
			\draw [fermion](a6)--(a10);
			\draw [fermion](a10)--(a7);
			\draw [vector](a10)--(a5);
			\draw [fermion](a5)--(a8);
			\draw [NR](a13)--(a12);
			\draw [vL](a12)--(a0);
			\draw [NR](a13)--(a14);
			\draw [vL](a14)--(a5);
			\node [left] at  (90:0.8) {$W_{L}$};
			\node [right] at  (a3) {$u_{L}$};
			\node [right] at  (a4) {$e_{L}$};
			\node [right] at  (a8) {$e_{L}$};
			\node [right] at  (a7) {$u_{L}$};
			\node [left] at  (a2) {$d_{L}$};
			\node [left] at  (a6) {$d_{L}$};
			\node [left] at  (0,-0.8) {$W_L$};
			\node at (0,0.33) {\textcolor{oorange}{$\bullet$}};
			\node [left] at  (0,0.45) {\textcolor{ppurple}{$\nu_{L}$}};
			\node [right] at  (0,0.18) {\textcolor{ggreen}{$N_{R}$}};
			\node [right] at  (0,-0.18) {\textcolor{ggreen}{$N_{R}$}};
                                \node [left] at  (0,-0.45) {\textcolor{ppurple}{$\nu_{L}$}};
			\node at (0,-0.33) {\textcolor{oorange}{$\bullet$}};
			\node at (0.005,0.0) {\textcolor{ggreen}{$\pmb{\times}$}};
			\node at (0,-1.35) {(a)};
		\end{tikzpicture}
		\begin{tikzpicture}[line width=1pt, scale=1.7,>=Stealth]
			\path (90:0.6) coordinate (a0);
			\path (90:1) coordinate (a1);
			\path (-0.8,1) coordinate (a2);
			\path (0.8,1) coordinate (a3);
			\path (0.8,0.6) coordinate (a4);
			\path (270:0.4) coordinate (a5);
			\path (-0.8,-1) coordinate (a6);
			\path (0.8,-1) coordinate (a7);
			\path (0.8,-0.4) coordinate (a8);
			\path (0.8,-1.4) coordinate (a9);
			\path (0,-1) coordinate (a10);
			\path (0,-1.4) coordinate (a11);
			\path (0,0.3) coordinate (a12);
			\path (0,-0.05) coordinate (a13);
			\path (0,-0.2) coordinate (a14);
			\draw [vector](a0)--(a1);
			\draw [fermion](a2)--(a1);
			\draw [fermion](a1)--(a3);
			\draw [fermion](a0)--(a4);
			\draw [fermion](a6)--(a10);
			\draw [fermion](a10)--(a7);
			\draw [vector](a10)--(a5);
			\draw [fermion](a5)--(a8);
			\draw [NR](a13)--(a12);
			\draw [vL](a12)--(a0);
			\draw [NR](a13)--(a5);
			\node [left] at  (90:0.8) {$W_{L}$};
			\node [right] at  (a3) {$u_{L}$};
			\node [right] at  (a4) {$e_{L}$};
			\node [right] at  (a8) {$e_{R}$};
			\node [right] at  (a7) {$u_{L}$};
			\node [left] at  (a2) {$d_{L}$};
			\node [left] at  (a6) {$d_{L}$};
			\node [right] at  (0,-0.55) {$W_{R}$};
			\node [left] at  (0,-0.9) {$W_{L}$};
			\node at (0,0.26) {\textcolor{oorange}{$\bullet$}};
			\node [left] at  (0,0.45) {\textcolor{ppurple}{$\nu_{L}$}};
			\node [right] at  (0,0.1) {\textcolor{ggreen}{$N_{R}$}};
			\node [right] at  (0,-0.22) {\textcolor{ggreen}{$N_{R}$}};
			\node at (0.005,-0.05) {\textcolor{ggreen}{$\pmb{\times}$}};
			\node at (0,-0.722) {\textcolor{mmagenta}{$\pmb{\btimes}$}};
			\node at (0,-1.35) {(b)};
		\end{tikzpicture}
		\begin{tikzpicture}[line width=1pt, scale=1.7,>=Stealth]
			\path (90:0.6) coordinate (a0);
			\path (90:1) coordinate (a1);
			\path (-0.8,1) coordinate (a2);
			\path (0.8,1) coordinate (a3);
			\path (0.8,0.6) coordinate (a4);
			\path (270:0.6) coordinate (a5);
			\path (-0.8,-1) coordinate (a6);
			\path (0.8,-1) coordinate (a7);
			\path (0.8,-0.6) coordinate (a8);
			\path (0.8,-1.4) coordinate (a9);
			\path (0,-1) coordinate (a10);
			\path (0,-1.4) coordinate (a11);
			\path (0,0.2) coordinate (a12);
			\path (0,-0.2) coordinate (a13);
			\draw [vector](a0)--(a1);
			\draw [fermion](a2)--(a1);
			\draw [fermion](a1)--(a3);
			\draw [fermion](a0)--(a4);
			\draw [fermion](a6)--(a10);
			\draw [fermion](a10)--(a7);
			\draw [vector](a10)--(a5);
			\draw [fermion](a5)--(a8);
			\draw [NR](a13)--(a12);
			\draw [vL](a12)--(a0);
			\draw [NR](a13)--(a5);
			\node [left] at  (90:0.8) {$W_{L}$};
			\node [right] at  (a3) {$u_{L}$};
			\node [right] at  (a4) {$e_{L}$};
			\node [right] at  (a8) {$e_{R}$};
			\node [right] at  (a7) {$u_{R}$};
			\node [left] at  (a2) {$d_{L}$};
			\node [left] at  (a6) {$d_{R}$};
			\node [right] at  (0,-0.83) {$W_R$};
			\node at (0,0.18) {\textcolor{oorange}{$\bullet$}};
			\node [left] at  (0,0.4) {\textcolor{ppurple}{$\nu_{L}$}};
			\node [right] at  (0,0) {\textcolor{ggreen}{$N_{R}$}};
			\node [right] at  (0,-0.4) {\textcolor{ggreen}{$N_{R}$}};
			\node at (0.005,-0.2) {\textcolor{ggreen}{$\pmb{\times}$}};
			\node at (0,-1.35) {(c)};
		\end{tikzpicture}
		\caption{The Feynman diagrams in Type-I dominance scenario of left-right symmetric model.}
		\label{LR-I}
\end{figure}
\begin{figure}[t]
\centering
\includegraphics[height=4.5cm]{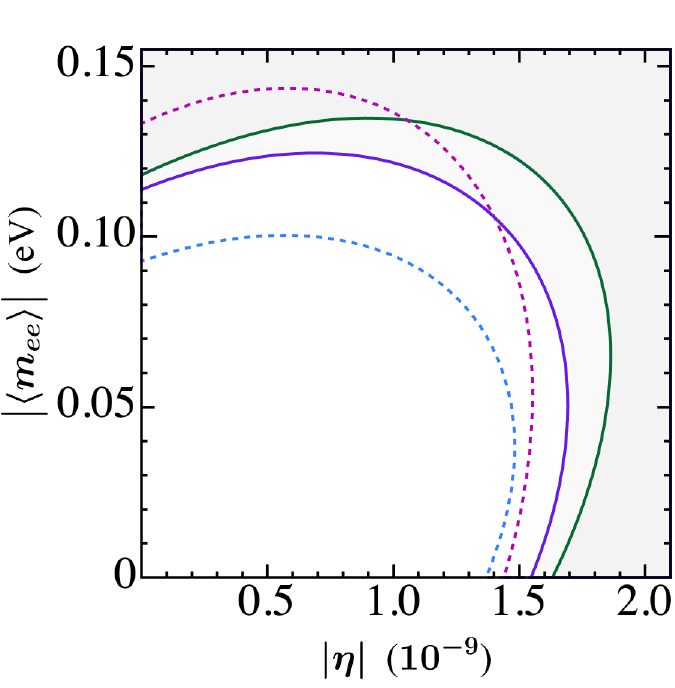}\quad
\includegraphics[height=4.5cm]{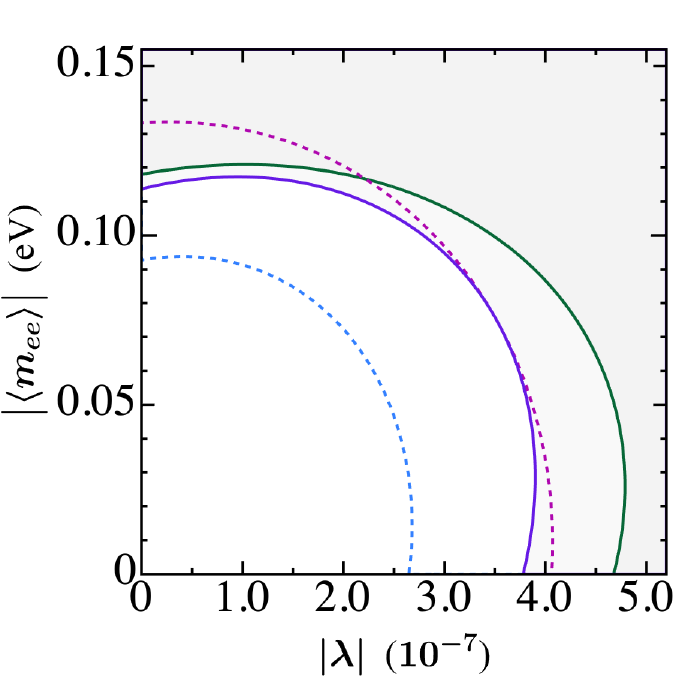}\quad
\includegraphics[height=4.5cm]{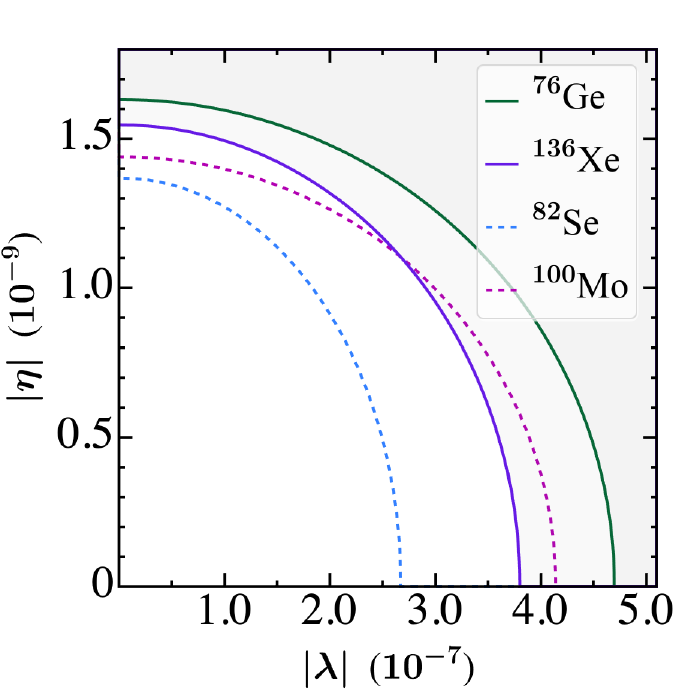}
\caption{These three figures show the relation between the effective light neutrino mass $\langle m_{ee}\rangle$ and the parameters $\eta,\lambda$, with the phase difference assumed to be zero. The solid green and purple lines correspond to the GERDA and KamLAND-Zen experimental results, and the gray regions are excluded. The dashed blue and magenta lines correspond to the isotopes $^{82}$Se and $^{100}$Mo with the half-life set to be $10^{26}$ yrs and $3\times10^{26}$ yrs, respectively.}\label{LR-I-plot}
\end{figure}

\paragraph{Type-I dominance}
If neutrino mass is generated through Type-I seesaw, which means the Majorana mass term $M_{\Delta_{L}}$ is negligible, the Dirac mass term $M_{\Phi\ell}$ and the Majorana mass term $M_{\Delta_{R}}$ contribute to $0\nu\beta\beta$ decay as the Fig.~\ref{LR-I} shown.
Besides the standard neutrino exchange, the leading contribution is from the long-range mechanism. The corresponding effective operators are 
\begin{align}
\mathcal{L}_{\text{eff}}=\dfrac{G_{F}V_{ud}}{\sqrt{2}}[4\epsilon_{V-A}^{V+A}(\overline{u_{L}}\gamma_{\mu}d_{L})(\overline{e_{R}}\gamma^{\mu}\nu_{L}^{c})+4\epsilon_{V+A}^{V+A}(\overline{u_{L}}\gamma_{\mu}d_{L})(\overline{e_{R}}\gamma^{\mu}\nu_{L}^{c})]+{\text{h.c.}}\,,
\end{align}
where 
\begin{align}
\eta=\epsilon_{V-A}^{V+A}U_{ei}=\tan\alpha\sum\limits_{i}T_{ei}^{*}U_{ei}\,,\quad \lambda=\epsilon_{V+A}^{V+A}U_{ei}=\dfrac{m_{W_L}^{2}}{m_{W_{R}}^{2}}\sum\limits_{i}T_{ei}^{*}U_{ei}\,,\label{1b1c}
\end{align}
are the dimensionless coefficients correspond to Fig.~\ref{LR-I} (b) and (c) with $\alpha$ be the mixing angle of $W_{L}$ and $W_{R}$. The $0\nu\beta\beta$ inverse half-life is derived as
\begin{align}
T_{1/2}^{-1}&=G_{11}^{(0)}\bigg|\dfrac{\langle m_{ee}\rangle}{m_{e}}\mathcal{M}_{\nu}\bigg|^{2}
+G_{33}^{(0)}\big|\eta\mathcal{M}_{3,3}^{LL}+\lambda\mathcal{M}_{3,3}^{LR}\big|^{2}
+G_{44}^{(0)}\big|\eta\mathcal{M}_{3,4}^{LL}+\lambda\mathcal{M}_{3,4}^{LR}\big|^{2}+G_{55}^{(0)}\big|\eta\mathcal{M}_{3,5}^{LR}\big|^{2}\notag\\
&+G_{66}^{(0)}\big|\eta\mathcal{M}_{3,6}^{LR}\big|^{2}
+2G_{15}^{(0)}{\text{Re}}\bigg[\dfrac{\langle m_{ee}\rangle}{m_{e}}\mathcal{M}_{\nu}\big(\eta\mathcal{M}_{3,5}^{LR}\big)^{*}\bigg]
+2G_{16}^{(0)}{\text{Re}}\bigg[\dfrac{\langle m_{ee}\rangle}{m_{e}}\mathcal{M}_{\nu}\big(\eta\mathcal{M}_{3,6}^{LR}\big)^{*}\bigg]
\notag\\
&+2G_{13}^{(0)}{\text{Re}}\bigg[\dfrac{\langle m_{ee}\rangle}{m_{e}}\mathcal{M}_{\nu}\big(\eta\mathcal{M}_{3,3}^{LL}+\lambda\mathcal{M}_{3,3}^{LR}\big)^{*}\bigg]
+2G_{14}^{(0)}{\text{Re}}\bigg[\dfrac{\langle m_{ee}\rangle}{m_{e}}\mathcal{M}_{\nu}\big(\eta\mathcal{M}_{3,4}^{LL}+\lambda\mathcal{M}_{3,4}^{LR}\big)^{*}\bigg]\notag\\
&+2G_{34}^{(0)}{\text{Re}}\big[\big(\eta\mathcal{M}_{3,3}^{LL}+\lambda\mathcal{M}_{3,3}^{LR}\big)\big(\eta\mathcal{M}_{3,4}^{LL}+\lambda\mathcal{M}_{3,4}^{LR}\big)^{*}\big]
+2G_{56}^{(0)}{\text{Re}}\bigg[\eta\mathcal{M}_{3,5}^{LR}\big(\eta\mathcal{M}_{3,6}^{LR}\big)^{*}\bigg]\,.
\end{align}
The numerical analysis takes the values of NMEs and PSFs from~\cite{Kotila:2021xgw}, and the phase difference between $\langle m_{ee}\rangle$ and $\eta,\lambda$ is set to be zero. We show the contours between the parameters $\eta,\lambda$ and effective neutrino mass $\langle m_{ee}\rangle$ in Fig.~\ref{LR-I-plot}. There are slight differences among these elliptical regions in different isotopes, and it is still possible to narrow the parameter space to some extent with the combination searches within multiple isotopes.

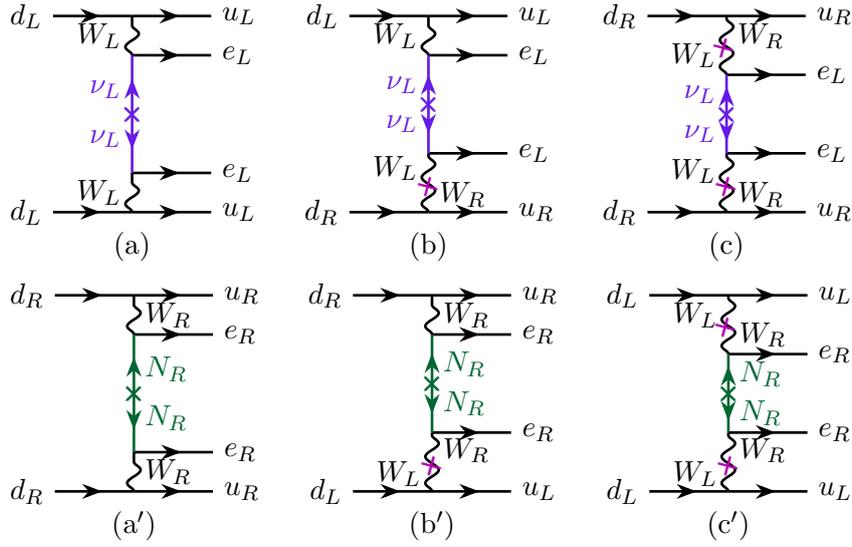
\begin{figure}[h]
\centering
\begin{tikzpicture}[line width=1pt, scale=1.3,>=Stealth]
			\path (90:0.6) coordinate (a0);
			\path (90:1) coordinate (a1);
			\path (-0.8,1) coordinate (a2);
			\path (0.8,1) coordinate (a3);
			\path (0.8,0.6) coordinate (a4);
			\path (270:0.6) coordinate (a5);
			\path (-0.8,-1) coordinate (a6);
			\path (0.8,-1) coordinate (a7);
			\path (0.8,-0.6) coordinate (a8);
			\path (0.8,-1.4) coordinate (a9);
			\path (0,-1) coordinate (a10);
			\path (0,-1.4) coordinate (a11);
			\path (0,0.4) coordinate (a12);
			\path (0,0.0) coordinate (a13);
			\path (0,-0.4) coordinate (a14);
			\draw [vector](a0)--(a1);
			\draw [fermion](a2)--(a1);
			\draw [fermion](a1)--(a3);
			\draw [fermion](a0)--(a4);
			\draw [fermion](a6)--(a10);
			\draw [fermion](a10)--(a7);
			\draw [vector](a10)--(a5);
			\draw [fermion](a5)--(a8);
			\draw [vL](a13)--(a0);
			\draw [vL](a13)--(a5);
			\node [left] at  (90:0.8) {$W_{L}$};
			\node [right] at  (a3) {$u_{L}$};
			\node [right] at  (a4) {$e_{L}$};
			\node [right] at  (a8) {$e_{L}$};
			\node [right] at  (a7) {$u_{L}$};
			\node [left] at  (a2) {$d_{L}$};
			\node [left] at  (a6) {$d_{L}$};
			\node [left] at  (0,-0.8) {$W_L$};
			\node [left] at  (0,0.25) {\textcolor{ppurple}{$\nu_{L}$}};
                                \node [left] at  (0,-0.25) {\textcolor{ppurple}{$\nu_{L}$}};
			\node at (0.005,0.0) {\textcolor{ppurple}{$\pmb{\times}$}};
			\node at (0,-1.35) {(a)};
		\end{tikzpicture}\quad
		\begin{tikzpicture}[line width=1pt, scale=1.3,>=Stealth]
			\path (90:0.6) coordinate (a0);
			\path (90:1) coordinate (a1);
			\path (-0.8,1) coordinate (a2);
			\path (0.8,1) coordinate (a3);
			\path (0.8,0.6) coordinate (a4);
			\path (270:0.4) coordinate (a5);
			\path (-0.8,-1) coordinate (a6);
			\path (0.8,-1) coordinate (a7);
			\path (0.8,-0.4) coordinate (a8);
			\path (0.8,-1.4) coordinate (a9);
			\path (0,-1) coordinate (a10);
			\path (0,-1.4) coordinate (a11);
			\path (0,0.1) coordinate (a12);
			\path (0,-0.05) coordinate (a13);
			\path (0,-0.2) coordinate (a14);
			\draw [vector](a0)--(a1);
			\draw [fermion](a2)--(a1);
			\draw [fermion](a1)--(a3);
			\draw [fermion](a0)--(a4);
			\draw [fermion](a6)--(a10);
			\draw [fermion](a10)--(a7);
			\draw [vector](a10)--(a5);
			\draw [fermion](a5)--(a8);
			\draw [vL](a12)--(a5);
			\draw [vL](a12)--(a0);
			\node [left] at  (90:0.8) {$W_{L}$};
			\node [right] at  (a3) {$u_{L}$};
			\node [right] at  (a4) {$e_{L}$};
			\node [right] at  (a8) {$e_{L}$};
			\node [right] at  (a7) {$u_{R}$};
			\node [left] at  (a2) {$d_{L}$};
			\node [left] at  (a6) {$d_{R}$};
			\node [left] at  (0,-0.55) {$W_{L}$};
			\node [right] at  (0,-0.82) {$W_{R}$};
			\node [left] at  (0,0.3) {\textcolor{ppurple}{$\nu_{L}$}};
			\node [left] at  (0,-0.1) {\textcolor{ppurple}{$\nu_{L}$}};
			\node at (0.005,0.1) {\textcolor{ppurple}{$\pmb{\times}$}};
			\node at (0,-0.735) {\textcolor{mmagenta}{$\pmb{\btimes}$}};
			\node at (0,-1.35) {(b)};
		\end{tikzpicture}\quad
		\begin{tikzpicture}[line width=1pt, scale=1.3,>=Stealth]
			\path (90:0.4) coordinate (a0);
			\path (90:1) coordinate (a1);
			\path (-0.8,1) coordinate (a2);
			\path (0.8,1) coordinate (a3);
			\path (0.8,0.4) coordinate (a4);
			\path (270:0.4) coordinate (a5);
			\path (-0.8,-1) coordinate (a6);
			\path (0.8,-1) coordinate (a7);
			\path (0.8,-0.4) coordinate (a8);
			\path (0.8,-1.4) coordinate (a9);
			\path (0,-1) coordinate (a10);
			\path (0,-1.4) coordinate (a11);
			\path (0,0.0) coordinate (a12);
			\draw [vector](a0)--(a1);
			\draw [fermion](a2)--(a1);
			\draw [fermion](a1)--(a3);
			\draw [fermion](a0)--(a4);
			\draw [fermion](a6)--(a10);
			\draw [fermion](a10)--(a7);
			\draw [vector](a10)--(a5);
			\draw [fermion](a5)--(a8);
			\draw [vLp](a12)--(a5);
			\draw [vLp](a12)--(a0);
			\node [left] at  (90:0.6) {$W_{L}$};
			\node [right] at  (90:0.8) {$W_{R}$};
			\node [right] at  (a3) {$u_{R}$};
			\node [right] at  (a4) {$e_{L}$};
			\node [right] at  (a8) {$e_{L}$};
			\node [right] at  (a7) {$u_{R}$};
			\node [left] at  (a2) {$d_{R}$};
			\node [left] at  (a6) {$d_{R}$};
			\node [left] at  (0,-0.58) {$W_{L}$};
			\node [right] at  (0,-0.82) {$W_{R}$};
			\node [left] at  (0,0.2) {\textcolor{ppurple}{$\nu_{L}$}};
			\node [left] at  (0,-0.2) {\textcolor{ppurple}{$\nu_{L}$}};
			\node at (0.005,0) {\textcolor{ppurple}{$\pmb{\times}$}};
			\node at (-0.025,0.68) {\textcolor{mmagenta}{$\pmb{\btimes}$}};
			\node at (0,-0.735) {\textcolor{mmagenta}{$\pmb{\btimes}$}};
			\node at (0,-1.35) {(c)};
		\end{tikzpicture}\\
		\begin{tikzpicture}[line width=1pt, scale=1.3,>=Stealth]
			\path (90:0.6) coordinate (a0);
			\path (90:1) coordinate (a1);
			\path (-0.8,1) coordinate (a2);
			\path (0.8,1) coordinate (a3);
			\path (0.8,0.6) coordinate (a4);
			\path (270:0.6) coordinate (a5);
			\path (-0.8,-1) coordinate (a6);
			\path (0.8,-1) coordinate (a7);
			\path (0.8,-0.6) coordinate (a8);
			\path (0.8,-1.4) coordinate (a9);
			\path (0,-1) coordinate (a10);
			\path (0,-1.4) coordinate (a11);
			\path (0,0.4) coordinate (a12);
			\path (0,0.0) coordinate (a13);
			\path (0,-0.4) coordinate (a14);
			\draw [vector](a0)--(a1);
			\draw [fermion](a2)--(a1);
			\draw [fermion](a1)--(a3);
			\draw [fermion](a0)--(a4);
			\draw [fermion](a6)--(a10);
			\draw [fermion](a10)--(a7);
			\draw [vector](a10)--(a5);
			\draw [fermion](a5)--(a8);
			\draw [NR](a13)--(a0);
			\draw [NR](a13)--(a5);
			\node [right] at  (90:0.8) {$W_{R}$};
			\node [right] at  (a3) {$u_{R}$};
			\node [right] at  (a4) {$e_{R}$};
			\node [right] at  (a8) {$e_{R}$};
			\node [right] at  (a7) {$u_{R}$};
			\node [left] at  (a2) {$d_{R}$};
			\node [left] at  (a6) {$d_{R}$};
			\node [right] at  (0,-0.8) {$W_R$};
			\node [right] at  (0,0.25) {\textcolor{ggreen}{$N_{R}$}};
                                \node [right] at  (0,-0.25) {\textcolor{ggreen}{$N_{R}$}};
			\node at (0.005,0.0) {\textcolor{ggreen}{$\pmb{\times}$}};
			\node at (0,-1.35) {(a$^{\prime}$)};
		\end{tikzpicture}\quad
		\begin{tikzpicture}[line width=1pt, scale=1.3,>=Stealth]
			\path (90:0.6) coordinate (a0);
			\path (90:1) coordinate (a1);
			\path (-0.8,1) coordinate (a2);
			\path (0.8,1) coordinate (a3);
			\path (0.8,0.6) coordinate (a4);
			\path (270:0.4) coordinate (a5);
			\path (-0.8,-1) coordinate (a6);
			\path (0.8,-1) coordinate (a7);
			\path (0.8,-0.4) coordinate (a8);
			\path (0.8,-1.4) coordinate (a9);
			\path (0,-1) coordinate (a10);
			\path (0,-1.4) coordinate (a11);
			\path (0,0.1) coordinate (a12);
			\path (0,-0.05) coordinate (a13);
			\path (0,-0.2) coordinate (a14);
			\draw [vector](a0)--(a1);
			\draw [fermion](a2)--(a1);
			\draw [fermion](a1)--(a3);
			\draw [fermion](a0)--(a4);
			\draw [fermion](a6)--(a10);
			\draw [fermion](a10)--(a7);
			\draw [vector](a10)--(a5);
			\draw [fermion](a5)--(a8);
			\draw [NR](a12)--(a5);
			\draw [NR](a12)--(a0);
			\node [right] at  (90:0.8) {$W_{R}$};
			\node [right] at  (a3) {$u_{R}$};
			\node [right] at  (a4) {$e_{R}$};
			\node [right] at  (a8) {$e_{R}$};
			\node [right] at  (a7) {$u_{L}$};
			\node [left] at  (a2) {$d_{R}$};
			\node [left] at  (a6) {$d_{L}$};
			\node [right] at  (0,-0.58) {$W_{R}$};
			\node [left] at  (0,-0.82) {$W_{L}$};
			\node [right] at  (0,0.3) {\textcolor{ggreen}{$N_{R}$}};
			\node [right] at  (0,-0.1) {\textcolor{ggreen}{$N_{R}$}};
			\node at (0.005,0.1) {\textcolor{ggreen}{$\pmb{\times}$}};
			\node at (0,-0.735) {\textcolor{mmagenta}{$\pmb{\btimes}$}};
			\node at (0,-1.35) {(b$^{\prime}$)};
		\end{tikzpicture}\quad
		\begin{tikzpicture}[line width=1pt, scale=1.3,>=Stealth]
			\path (90:0.4) coordinate (a0);
			\path (90:1) coordinate (a1);
			\path (-0.8,1) coordinate (a2);
			\path (0.8,1) coordinate (a3);
			\path (0.8,0.4) coordinate (a4);
			\path (270:0.4) coordinate (a5);
			\path (-0.8,-1) coordinate (a6);
			\path (0.8,-1) coordinate (a7);
			\path (0.8,-0.4) coordinate (a8);
			\path (0.8,-1.4) coordinate (a9);
			\path (0,-1) coordinate (a10);
			\path (0,-1.4) coordinate (a11);
			\path (0,0.0) coordinate (a12);
			\draw [vector](a0)--(a1);
			\draw [fermion](a2)--(a1);
			\draw [fermion](a1)--(a3);
			\draw [fermion](a0)--(a4);
			\draw [fermion](a6)--(a10);
			\draw [fermion](a10)--(a7);
			\draw [vector](a10)--(a5);
			\draw [fermion](a5)--(a8);
			\draw [NRp](a12)--(a5);
			\draw [NRp](a12)--(a0);
			\node [right] at  (90:0.6) {$W_{R}$};
			\node [left] at  (90:0.8) {$W_{L}$};
			\node [right] at  (a3) {$u_{L}$};
			\node [right] at  (a4) {$e_{R}$};
			\node [right] at  (a8) {$e_{R}$};
			\node [right] at  (a7) {$u_{L}$};
			\node [left] at  (a2) {$d_{L}$};
			\node [left] at  (a6) {$d_{L}$};
			\node [right] at  (0,-0.58) {$W_{R}$};
			\node [left] at  (0,-0.82) {$W_{L}$};
			\node [right] at  (0,0.2) {\textcolor{ggreen}{$N_{R}$}};
			\node [right] at  (0,-0.2) {\textcolor{ggreen}{$N_{R}$}};
			\node at (0.005,0) {\textcolor{ggreen}{$\pmb{\times}$}};
			\node at (-0.025,0.68) {\textcolor{mmagenta}{$\pmb{\btimes}$}};
			\node at (0,-0.735) {\textcolor{mmagenta}{$\pmb{\btimes}$}};
			\node at (0,-1.35) {(c$^{\prime}$)};
		\end{tikzpicture}
		\caption{The Feynman diagrams in Type-II dominance scenario of left-right symmetry model.}
		\label{LR-II}
\end{figure}

\paragraph{Type-II dominance}
In the Type-II dominance scenario, the Dirac mass term $M_{\Phi\ell}$ is negligible, and the tiny neutrino mass
is generated dominantly through the Majorana mass term $M_{\Delta_{L}}$. The Feynman diagrams of $0\nu\beta\beta$ decay are shown in Fig.~\ref{LR-II}.
The (a$^{\prime}$), (b$^{\prime}$), (c$^{\prime}$) Feynman diagrams are contributed by the Majorana mass term of $N_{R}$, while (b) and (c) Feynman diagrams by the Majorana mass term of $\nu_{L}$. The corresponding dimensionless coefficients are
\begin{gather}
\epsilon_{2a}=\dfrac{\langle m_{ee}\rangle}{m_{e}}\,,\quad
\epsilon_{2a^{\prime}}=\epsilon_{3}^{RRR}=\dfrac{m_{W_{L}}^{4}}{m_{W_{R}}^{4}}\sum\limits_{i}S_{ei}^{2}\dfrac{m_{p}}{m_{N_{i}}}\,,\notag\\
\epsilon_{2b}=\epsilon_{V+A}^{V-A}=\tan{\alpha}\dfrac{\langle m_{ee}\rangle}{m_{e}}\,,\quad
\epsilon_{2b^{\prime}}=\epsilon_{3}^{RLR}=\tan{\alpha}\dfrac{m_{W_{L}}^{4}}{m_{W_{R}}^{4}}\sum\limits_{i}S_{ei}^{2}\dfrac{m_{p}}{m_{N_{i}}}\,,\notag\\
\epsilon_{2c}=\tan^{2}{\alpha}\dfrac{\langle m_{ee}\rangle}{m_{e}}\,,\quad
\epsilon_{2c^{\prime}}=\epsilon_{3}^{LLR}=\tan^{2}{\alpha}\dfrac{m_{W_{L}}^{4}}{m_{W_{R}}^{4}}\sum\limits_{i}S_{ei}^{2}\dfrac{m_{p}}{m_{N_{i}}}\,.
\end{gather}

The current measurements give the mass of right-handed gauge boson $W_{R}$ to be heavier than 5~TeV~\cite{ATLAS:2018dcj,CMS:2018agk}, especially extends to 6.4~TeV for Majorana $N_{R}$ at $m_{N_{R}}<1$~TeV~\cite{ATLAS:2023cjo}. In our analysis, $m_{W_{R}}=7$~TeV is taken. The mixing between the left-handed and right-handed gauge boson could be neglected so that the dominant contributions are (a) and (a$^{\prime}$), which correspond to standard light neutrino exchange and right-handed heavy neutrino exchange, respectively. 
\begin{figure}[b]
\centering
\includegraphics[height=4.5cm]{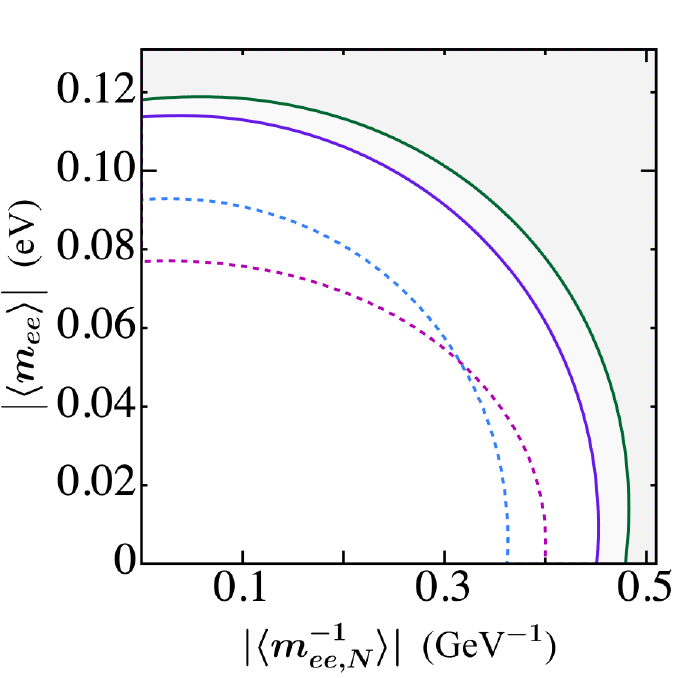}\qquad
\includegraphics[height=4.5cm]{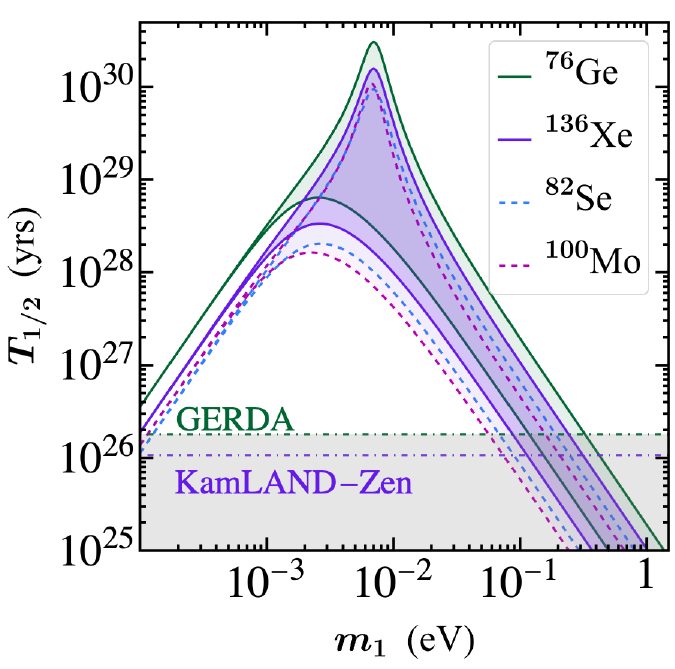}
\caption{The left one shows the relation between the effective light neutrino mass $\langle m_{ee}\rangle$ and the effective heavy neutrino mass $\langle m_{ee,N}^{-1}\rangle$ with the phase difference assumed to be zero. The mass of $SU(2)_{R}$ boson is fixed as $m_{W_{R}}=7~$TeV. The solid green and purple lines correspond to the GERDA and KamLAND-Zen experimental results, and the gray regions are excluded. The dashed blue and magenta lines correspond to the isotopes $^{82}$Se and $^{100}$Mo with the half-life set to be $10^{26}$ yrs. The right one shows the half-life of isotopes as a function of the lightest mass $m_{1}$ in the normal ordering.}
\label{LR-II-plot}
\end{figure}

The relation between the effective light neutrino mass $\langle m_{ee}\rangle$ and inverse effective heavy neutrino mass $\langle m_{ee,N}^{-1}\rangle\equiv\sum_{i}S_{ei}^{2}/m_{N_{i}}$ are shown in the left one in Fig.~\ref{LR-II-plot} with the phase difference is set to be zero. As there is a relation that $m_{4,5}=m_{1,2}m_{6}/m_{3}$, one can simplify the half-life into the function of the lightest active neutrino mass $m_{1}$~\cite{deVries:2022nyh}. The half-life of different isotopes is shown in the right one in Fig.~\ref{LR-II-plot}, assuming that the heaviest right-handed neutrino mass is 1 TeV. The combination of experimental searches in different isotopes can help to examine the parameter region.

\subsection{One-Loop Models with Leptoquarks}
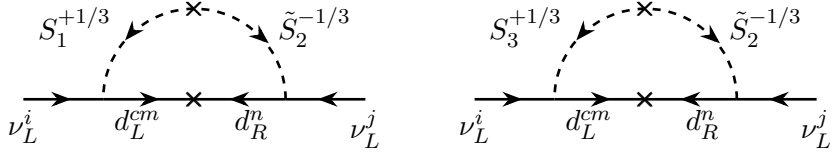
\begin{figure}[t]
	\begin{center}
		\begin{tikzpicture}[line width=1pt, scale=1.5,>=Stealth]
			\path (0:0) coordinate (a00);
			\path (180:0.8) coordinate (a0);
			\path (0:0.8) coordinate (a1);
			\path (180:1.5) coordinate (a2);
			\path (0:1.5) coordinate (a3);
			\path (90:0.8) coordinate (a4);
			\draw [fermionbar](a00)--(a0);
			\draw [fermionbar](a00)--(a1);
			\draw [fermion](a3)--(a1);
			\draw [fermion](a2)--(a0);
			\draw [scalar](a4) arc [start angle=90, end angle=180, radius=0.8];
			\draw [scalarbar](a1) arc [start angle=0, end angle=90, radius=0.8];
			\node [below] at  (0:0.5) {$d_{R}^{n}$};
			\node [below] at  (180:0.5) {$d_{L}^{cm}$};
			\node [below] at  (0:1.5) {$\nu_{L}^{j}$};
			\node [below] at  (180:1.5) {$\nu_{L}^{i}$};
			\node [left] at  (135:0.9) {$S_{1}^{+1/3}$};
			\node [right] at  (45:0.9) {$\tilde{S}_{2}^{-1/3}$};
			\node at  (a00) {$\pmb{\times}$};
			\node at  (90:0.8) {$\pmb{\times}$};
		\end{tikzpicture}\quad\quad
		\begin{tikzpicture}[line width=1pt, scale=1.5,>=Stealth]
			\path (0:0) coordinate (a00);
			\path (180:0.8) coordinate (a0);
			\path (0:0.8) coordinate (a1);
			\path (180:1.5) coordinate (a2);
			\path (0:1.5) coordinate (a3);
			\path (90:0.8) coordinate (a4);
			\draw [fermionbar](a00)--(a0);
			\draw [fermionbar](a00)--(a1);
			\draw [fermion](a3)--(a1);
			\draw [fermion](a2)--(a0);
			\draw [scalar](a4) arc [start angle=90, end angle=180, radius=0.8];
			\draw [scalarbar](a1) arc [start angle=0, end angle=90, radius=0.8];
			\node [below] at  (0:0.5) {$d_{R}^{n}$};
			\node [below] at  (180:0.5) {$d_{L}^{cm}$};
			\node [below] at  (0:1.5) {$\nu_{L}^{j}$};
			\node [below] at  (180:1.5) {$\nu_{L}^{i}$};
			\node [left] at  (135:0.9) {$S_{3}^{+1/3}$};
			\node [right] at  (45:0.9) {$\tilde{S}_{2}^{-1/3}$};
			\node at  (a00) {$\pmb{\times}$};
			\node at  (90:0.8) {$\pmb{\times}$};
		\end{tikzpicture}
	\caption{The one-loop level neutrino mass diagrams with $S_{1}\&\tilde{S}_{2}$ (left one) and $S_{3}\&\tilde{S}_{2}$ (right one).}
	\label{sLQnumass}
	\end{center}
\end{figure}

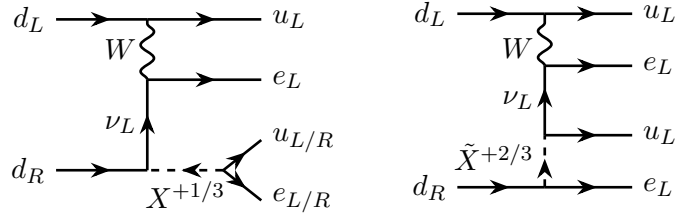
\begin{figure}[t]
	\begin{center}
		\begin{tikzpicture}[line width=1pt, scale=1,>=Stealth]
			\path (90:0.2) coordinate (a0);
			\path (90:1) coordinate (a1);
			\path (-1.2,1) coordinate (a2);
			\path (1.5,1) coordinate (a3);
			\path (1.5,0.2) coordinate (a4);
			\path (270:1) coordinate (a5);
			\path (-1.2,-1) coordinate (a6);
			\path (1,-1) coordinate (a7);
			\path (1.5,-0.6) coordinate (a8);
			\path (1.5,-1.4) coordinate (a9);
			\draw [vector](a0)--(a1);
			\draw [fermion](a2)--(a1);
			\draw [fermion](a1)--(a3);
			\draw [fermion](a0)--(a4);
			\draw [fermionbar](a0)--(a5);
			\draw [fermion](a6)--(a5);
			\draw [scalar](a7)--(a5);
			\draw [fermion](a7)--(a8);
			\draw [fermion](a7)--(a9);
			\node [left] at  (90:0.6) {$W$};
			\node [right] at  (a3) {$u_{L}$};
			\node [right] at  (a4) {$e_{L}$};
			\node [right] at  (a8) {$u_{L/R}$};
			\node [right] at  (a9) {$e_{L/R}$};
			\node [left] at  (a2) {$d_{L}$};
			\node [left] at  (a6) {$d_{R}$};
			\node [left] at  (0,-0.4) {$\nu_{L}$};
			\node [below] at  (0.5,-1) {$X^{+1/3}$};
		\end{tikzpicture}\quad\quad
		\begin{tikzpicture}[line width=1pt, scale=1.15,>=Stealth]
			\path (90:0.4) coordinate (a0);
			\path (90:1) coordinate (a1);
			\path (-1,1) coordinate (a2);
			\path (1,1) coordinate (a3);
			\path (1,0.4) coordinate (a4);
			\path (270:0.4) coordinate (a5);
			\path (-1,-1) coordinate (a6);
			\path (1,-1) coordinate (a7);
			\path (1,-0.4) coordinate (a8);
			\path (1,-1.4) coordinate (a9);
			\path (0,-1) coordinate (a10);
			\path (0,-1.4) coordinate (a11);
			\draw [vector](a0)--(a1);
			\draw [fermion](a2)--(a1);
			\draw [fermion](a1)--(a3);
			\draw [fermion](a0)--(a4);
			\draw [fermionbar](a0)--(a5);
			\draw [fermion](a6)--(a10);
			\draw [fermion](a10)--(a7);
			\draw [scalar](a10)--(a5);
			\draw [fermion](a5)--(a8);
			\node [left] at  (90:0.6) {$W$};
			\node [right] at  (a3) {$u_{L}$};
			\node [right] at  (a4) {$e_{L}$};
			\node [right] at  (a8) {$u_{L}$};
			\node [right] at  (a7) {$e_{L}$};
			\node [left] at  (a2) {$d_{L}$};
			\node [left] at  (a6) {$d_{R}$};
			\node [left] at  (0,0) {$\nu_{L}$};
			\node [left] at  (0,-0.65) {$\tilde{X}^{+2/3}$};
		\end{tikzpicture}
	\caption{The contributions to neutrinoless double beta decay from the scalar components with charge $-1/3$ and $+2/3$. The $S_1\&\tilde{S}_{2}$ case contributes to the left one while the $S_{3}\&\tilde{S}_{2}$ case contributes to both two diagrams.}
	\label{LQ0vbb}
	\end{center}
\end{figure}

The models that contain leptoquarks have been discussed in some previous literature, e.g.~\cite{Crivellin:2017zlb,Becirevic:2018afm,Dorsner:2019itg,Babu:2020hun,Saad:2020ucl,Dorsner:2020aaz,DaRold:2020bib,Nomura:2021oeu,Zhang:2021dgl,Becirevic:2022tsj,Chen:2022hle}. Here, we consider the cases that can generate neutrino mass through the one-loop level with the internal scalar particles as leptoquarks. There are two cases, $S_{1}\&\tilde{S}_{2}$ and $S_{3}\&\tilde{S}_{2}$, that can generate neutrino mass as shown in Fig.~\ref{sLQnumass}.%
~%
The contributions of leptoquarks to the neutrinoless double beta decay have generally been discussed in~\cite{Hirsch:1996ye}, where the contributions are treated via the long-range mechanism. The Feynman diagrams of $0\nu\beta\beta$ decay are shown in Fig.~\ref{LQ0vbb}.
 In the following discussion, the first case is focused. The Yukawa interactions of $S_{1}=S_{1}^{+1/3}\sim(\bar{3},1,1/3)$ and $\tilde{S}_{2}=(\tilde{S}_{2}^{+2/3},\tilde{S}_{2}^{-1/3})^{T}$ $\sim(3,2,1/6)$ leptoquarks in fermion mass eigenstates are
\begin{align}
	\mathcal{L}_{Y}=&-y_{1RR}^{ij}\overline{u_{R}^{ci}}e_{R}^{j}S_{1}^{+1/3}
	-(V^{*}y_{1LL})^{ij}\overline{u_{L}^{ci}}e_{L}^{j}S_{1}^{+1/3}
	+(y_{1LL}U)^{ij}\overline{d_{L}^{ci}}\nu_{L}^{j}S_{1}^{+1/3}\notag\\
	&-\tilde{y}_{2RL}^{ij}\overline{d_{R}^{i}}\tilde{S}_{2}^{+2/3} e_{L}^{j}
	+(\tilde{y}_{2RL}U)^{ij}\overline{d_{R}^{i}}\tilde{S}_{2}^{-1/3}\nu_{L}^{j}+\text{h.c.}\,,
\end{align}
where $U$ is the Pontecorvo-Maki-Nakagawa-Sakata (PMNS) matrix and $V$ is the Cabibbo-Kobayashi-Maskawa (CKM) matrix. The long-range effective operators are
\begin{align}
\mathcal{L}_{\text{eff}}=\dfrac{G_{F}V_{ud}}{\sqrt{2}}\big[&4\epsilon_{S+P}^{S+P}(\overline{u_{L}}d_{R})(\overline{e_{L}}\nu_{L}^{c})+4\epsilon_{T_{R}}^{T_{R}}(\overline{u_{L}}\sigma_{\mu\nu}d_{R})(\overline{e_{L}}\sigma^{\mu\nu}\nu_{L}^{c})\notag\\
&+4\epsilon_{V+A}^{V+A}(\overline{u_{R}}\gamma_{\mu}d_{R})(\overline{e_{R}}\gamma^{\mu}\nu_{L}^{c})\big]+\text{h.c.}\,,
\end{align}
with the Wilson coefficients to be
\begin{align}
&\epsilon_{S+P}^{S+P}=\dfrac{1}{4\sqrt{2}G_{F}}\cos{\varphi}\sin{\varphi}\tilde{y}^{\prime *11}_{2RL}y_{1LL}^{\prime *11}\bigg(\dfrac{1}{M_{X_{1}}^{2}}-\dfrac{1}{M_{X_{2}}^{2}}\bigg)\,,\quad \epsilon_{T_{R}}^{T_{R}}=\dfrac{1}{4}\epsilon_{S+P}^{S+P}\,,\\
&\epsilon_{V+A}^{V+A}=\dfrac{1}{4\sqrt{2}G_{F}}\cos{\varphi}\sin{\varphi}\tilde{y}^{\prime *11}_{2RL}y_{1RR}^{*11}\bigg(\dfrac{1}{M_{X_{1}}^{2}}-\dfrac{1}{M_{X_{2}}^{2}}\bigg)\,,
\end{align}
where $\tilde{y}_{2RL}^{\prime 11}=(\tilde{y}_{2RL}U)^{11}$, $y_{1LL}^{\prime 11}=(V^{*}\tilde{y}_{1LL})^{11}$, and the $X_{1,2}$ are the mass eigenstates of $S_{1,2}^{+1/3}$. The angle $\varphi$ describes the mixing between $S_{1}^{+1/3}$ and $\tilde{S}_{2}^{+1/3}$, $\tan{2\varphi}=\sqrt{2}\mu v/(m_{S_{1}}^{2}-m_{\tilde{S}_{2}}^{2})$ with $\mu$ from the trilinear term of scalar potential $\mu S_{1}^{-1/3}\phi^{0}\tilde{S}_{2}^{+1/3}$ and $v$ to be the vacuum expectation value $\langle\phi^{0}\rangle=v/\sqrt{2}$.
\begin{figure}[t]
\centering
\includegraphics[height=4.5cm]{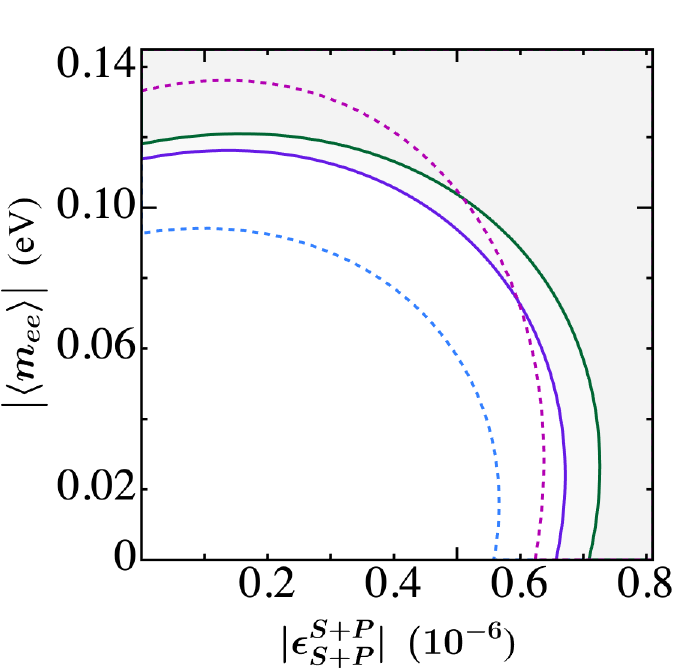}\quad
\includegraphics[height=4.5cm]{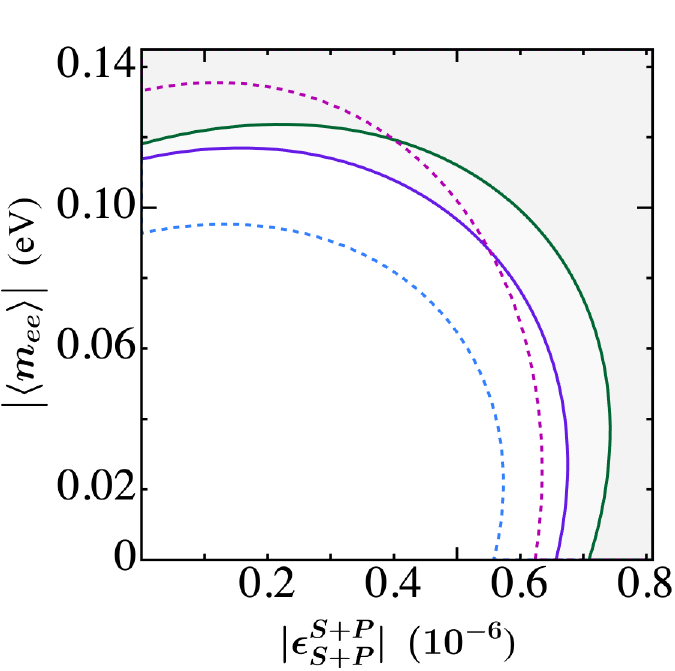}\quad
\includegraphics[height=4.5cm]{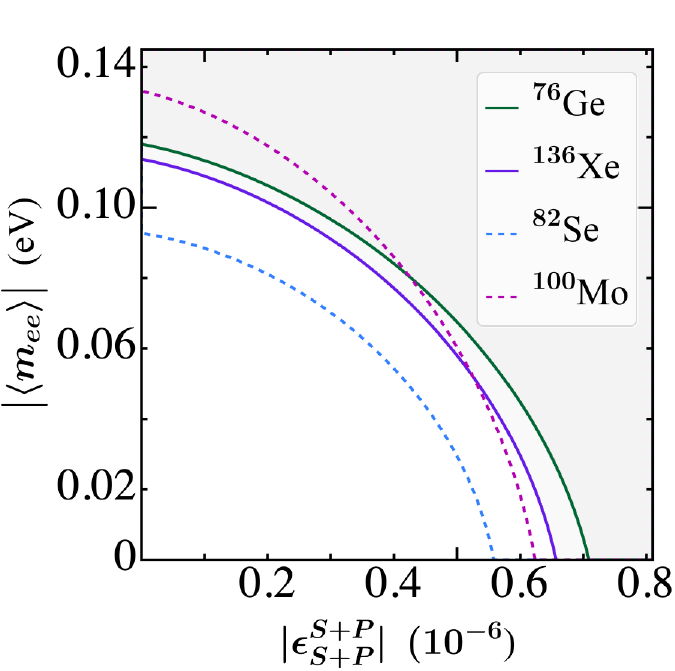}\quad
\caption{The relation between the effective neutrino mass $\langle m_{ee}\rangle$ and the parameter $\epsilon_{S+P}^{S+P}$ with the phase difference taken to be $\pi$ (left), $\pi/2$ (middle) and 0 (right). The solid green and purple lines correspond to the GERDA and KamLAND-Zen experimental results, and the gray regions are excluded. The dashed blue and magenta lines correspond to the isotopes $^{82}$Se and $^{100}$Mo with the half-life set to be $10^{26}$ yrs and $3\times10^{26}$ yrs.}\label{LQplot}
\end{figure}
The parameter $y_{1RR}^{11}$ is set as $y_{1RR}^{11}=0$ to simplify the numerical calculation so that the inverse half-life can be written as
\begin{align}
T_{1/2}^{-1}=&G_{11}^{(0)}\bigg|\dfrac{\langle m_{ee}\rangle}{m_{e}}\mathcal{M}_{\nu}\bigg|^{2}
+G_{11}^{\prime(0)}\big|\epsilon_{T_{R}}^{T_{R}}\mathcal{M}_{4,1}^{LR}+\epsilon_{S+P}^{S+P}\mathcal{M}_{5,1}^{LR}\big|^{2}\notag\\
&+2G_{11}^{\prime\prime(0)}{\text{Re}}\bigg[\dfrac{\langle m_{ee}\rangle}{m_{e}}\mathcal{M}_{\nu}(\epsilon_{T_{R}}^{T_{R}}\mathcal{M}_{4,1}^{LR}+\epsilon_{S+P}^{S+P}\mathcal{M}_{5,1}^{LR})^{*}\bigg]\,.
\end{align}
Substituting $\epsilon_{T_{R}}^{T_{R}}$ by $\epsilon_{S+P}^{S+P}/4$, one can get the correlation between the effective neutrino mass and the parameter $\epsilon_{S+P}^{S+P}$ as shown in Fig.~\ref{LQplot}. The figure shows the survival region when the phase difference is taken to be $\pi$ (left), $\pi/2$ (middle), and 0 (right), where one could find that the effects of interference are not obvious. The tilt angle in $^{100}\text{Mo}$ has a slight difference which indicates that a more comprehensive examination of the parameter space could be achieved by experiments within multiple isotopes.

\subsection{R-parity violating Supersymmetry Model}
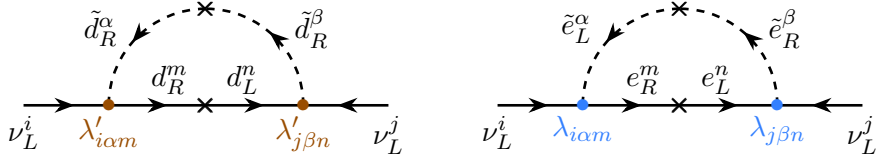
\begin{figure}[t]
	\begin{center}
		\begin{tikzpicture}[line width=1pt, scale=1.6,>=Stealth]
			\path (0:0) coordinate (a00);
			\path (180:0.8) coordinate (a0);
			\path (0:0.8) coordinate (a1);
			\path (180:1.5) coordinate (a2);
			\path (0:1.5) coordinate (a3);
			\path (90:0.8) coordinate (a4);
			\draw [fermionbar](a00)--(a0);
			\draw [fermion](a00)--(a1);
			\draw [fermion](a3)--(a1);
			\draw [fermion](a2)--(a0);
			\draw [scalar](a4) arc [start angle=90, end angle=180, radius=0.8];
			\draw [scalar](a1) arc [start angle=0, end angle=90, radius=0.8];
			\node [above] at  (0:0.3) {$d_{L}^{n}$};
			\node [above] at  (180:0.3) {$d_{R}^{m}$};
			\node [below] at  (1.5,0) {$\nu_{L}^{j}$};
			\node [below] at  (-1.5,0) {$\nu_{L}^{i}$};
			\node [left] at  (135:0.9) {$\tilde{d}_{R}^{\alpha}$};
			\node [right] at  (45:0.9) {$\tilde{d}_{R}^{\beta}$};
			\node at  (a00) {$\pmb{\times}$};
			\node at  (90:0.8) {$\pmb{\times}$};
			\node at  (0.8,0) {$\textcolor{bbrown}{\bullet}$};
			\node at  (-0.8,0) {$\textcolor{bbrown}{\bullet}$};
			\node [below] at  (-0.8,0) {\textcolor{bbrown}{$\lambda_{i\alpha m}^{\prime}$}};
			\node [below] at  (0.8,0) {\textcolor{bbrown}{$\lambda_{j\beta n}^{\prime}$}};
		\end{tikzpicture}\quad\quad
		\begin{tikzpicture}[line width=1pt, scale=1.6,>=Stealth]
			\path (0:0) coordinate (a00);
			\path (180:0.8) coordinate (a0);
			\path (0:0.8) coordinate (a1);
			\path (180:1.5) coordinate (a2);
			\path (0:1.5) coordinate (a3);
			\path (90:0.8) coordinate (a4);
			\draw [fermionbar](a00)--(a0);
			\draw [fermion](a00)--(a1);
			\draw [fermion](a3)--(a1);
			\draw [fermion](a2)--(a0);
			\draw [scalar](a4) arc [start angle=90, end angle=180, radius=0.8];
			\draw [scalar](a1) arc [start angle=0, end angle=90, radius=0.8];
			\node [above] at  (0:0.3) {$e_{L}^{n}$};
			\node [above] at  (180:0.3) {$e_{R}^{m}$};
			\node [below] at  (0:1.5) {$\nu_{L}^{j}$};
			\node [below] at  (180:1.5) {$\nu_{L}^{i}$};
			\node [left] at  (135:0.9) {$\tilde{e}_{L}^{\alpha}$};
			\node [right] at  (45:0.9) {$\tilde{e}_{R}^{\beta}$};
			\node at  (a00) {$\pmb{\times}$};
			\node at  (90:0.8) {$\pmb{\times}$};
			\node at  (0.8,0) {$\textcolor{bblue}{\bullet}$};
			\node at  (-0.8,0) {$\textcolor{bblue}{\bullet}$};
			\node [below] at  (-0.8,0) {\textcolor{bblue}{$\lambda_{i\alpha m}$}};
			\node [below] at  (0.8,0) {\textcolor{bblue}{$\lambda_{j\beta n}$}};
		\end{tikzpicture}
	\caption{The one-loop level neutrino mass diagrams with squark (lepton) and slepton (right one) mediator.}
	\label{SUSY-mv}
	\end{center}
\end{figure}

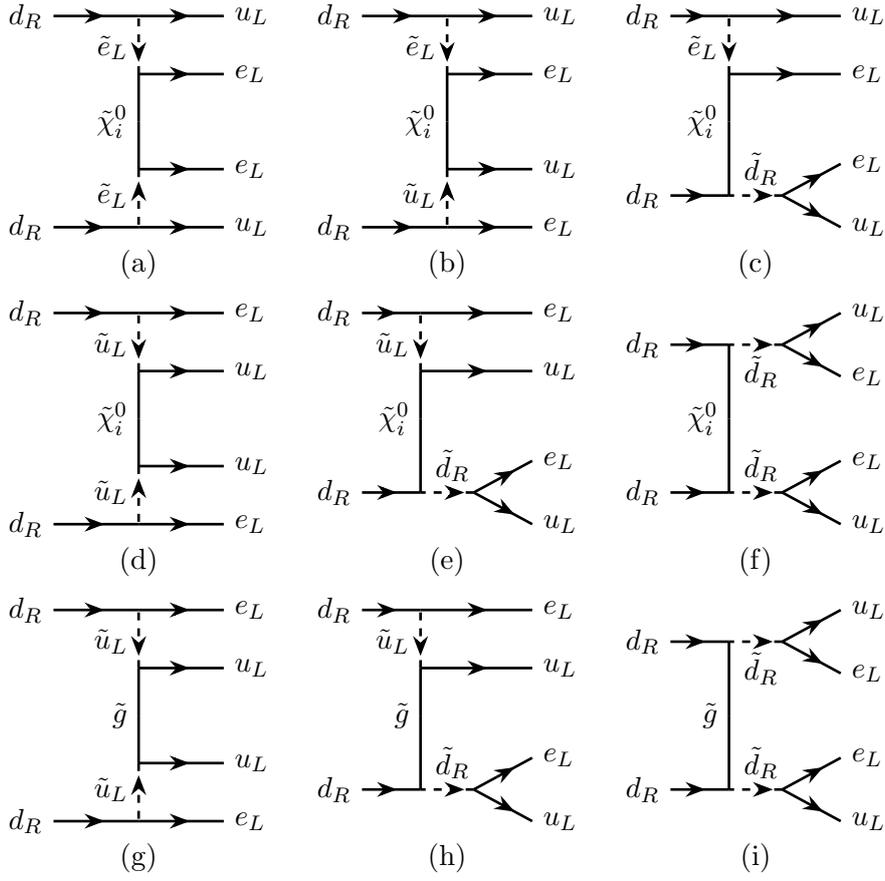
\begin{figure}[t]
\centering
\begin{tikzpicture}[line width=1pt, scale=1.4,>=Stealth]
			\path (90:0.45) coordinate (a0);
			\path (90:1) coordinate (a1);
			\path (-0.8,1) coordinate (a2);
			\path (0.8,1) coordinate (a3);
			\path (0.8,0.45) coordinate (a4);
			\path (270:0.45) coordinate (a5);
			\path (-0.8,-1) coordinate (a6);
			\path (0.8,-1) coordinate (a7);
			\path (0.8,-0.45) coordinate (a8);
			\path (0.8,-1.4) coordinate (a9);
			\path (0,-1) coordinate (a10);
			\path (0,-1.4) coordinate (a11);
			\path (0,0.4) coordinate (a12);
			\path (0,0.0) coordinate (a13);
			\path (0,-0.4) coordinate (a14);
			\draw [scalarbar](a0)--(a1);
			\draw [fermion](a2)--(a1);
			\draw [fermion](a1)--(a3);
			\draw [fermion](a0)--(a4);
			\draw [fermion](a6)--(a10);
			\draw [fermion](a10)--(a7);
			\draw [scalarbar](a5)--(a10);
			\draw [fermion](a5)--(a8);
			\draw [fermionnoarrow](a13)--(a0);
			\draw [fermionnoarrow](a13)--(a5);
			\node [left] at  (90:0.7) {$\tilde{e}_{L}$};
			\node [right] at  (a3) {$u_{L}$};
			\node [right] at  (a4) {$e_{L}$};
			\node [right] at  (a8) {$e_{L}$};
			\node [right] at  (a7) {$u_{L}$};
			\node [left] at  (a2) {$d_{R}$};
			\node [left] at  (a6) {$d_{R}$};
			\node [left] at  (0,-0.7) {$\tilde{e}_{L}$};
			\node [left] at  (0,0) {$\tilde{\chi}_{i}^{0}$};
			\node at (0,-1.35) {(a)};
		\end{tikzpicture}\quad
		\begin{tikzpicture}[line width=1pt, scale=1.4,>=Stealth]
			\path (90:0.45) coordinate (a0);
			\path (90:1) coordinate (a1);
			\path (-0.8,1) coordinate (a2);
			\path (0.8,1) coordinate (a3);
			\path (0.8,0.45) coordinate (a4);
			\path (270:0.45) coordinate (a5);
			\path (-0.8,-1) coordinate (a6);
			\path (0.8,-1) coordinate (a7);
			\path (0.8,-0.45) coordinate (a8);
			\path (0.8,-1.4) coordinate (a9);
			\path (0,-1) coordinate (a10);
			\path (0,-1.4) coordinate (a11);
			\path (0,0.4) coordinate (a12);
			\path (0,0.0) coordinate (a13);
			\path (0,-0.4) coordinate (a14);
			\draw [scalarbar](a0)--(a1);
			\draw [fermion](a2)--(a1);
			\draw [fermion](a1)--(a3);
			\draw [fermion](a0)--(a4);
			\draw [fermion](a6)--(a10);
			\draw [fermion](a10)--(a7);
			\draw [scalarbar](a5)--(a10);
			\draw [fermion](a5)--(a8);
			\draw [fermionnoarrow](a13)--(a0);
			\draw [fermionnoarrow](a13)--(a5);
			\node [left] at  (90:0.7) {$\tilde{e}_{L}$};
			\node [right] at  (a3) {$u_{L}$};
			\node [right] at  (a4) {$e_{L}$};
			\node [right] at  (a8) {$u_{L}$};
			\node [right] at  (a7) {$e_{L}$};
			\node [left] at  (a2) {$d_{R}$};
			\node [left] at  (a6) {$d_{R}$};
			\node [left] at  (0,-0.7) {$\tilde{u}_{L}$};
			\node [left] at  (0,0) {$\tilde{\chi}_{i}^{0}$};
			\node at (0,-1.35) {(b)};
		\end{tikzpicture}\quad
		\begin{tikzpicture}[line width=1pt, scale=1.4,>=Stealth]
			\path (-0.25,0.45) coordinate (a0);
			\path (-0.25,1) coordinate (a1);
			\path (-0.8,1) coordinate (a2);
			\path (0.8,1) coordinate (a3);
			\path (0.8,0.45) coordinate (a4);
			\path (-0.25,-0.7) coordinate (a5);
			\path (-0.8,-0.7) coordinate (a6);
			\path (0.8,-1) coordinate (a7);
			\path (-0.25,0.0) coordinate (a13);
			\draw [scalarbar](a0)--(a1);
			\draw [fermion](a2)--(a1);
			\draw [fermion](a1)--(a3);
			\draw [fermion](a0)--(a4);
			\draw [fermion](-0.8,-0.7)--(-0.25,-0.7);
			\draw [scalarbar](0.25,-0.7)--(-0.25,-0.7);
			\draw [fermionbar](0.8,-1)--(0.25,-0.7);
			\draw [fermionbar](0.8,-0.4)--(0.25,-0.7);
			\draw [fermionnoarrow](a13)--(a0);
			\draw [fermionnoarrow](a13)--(a5);
			\node [left] at  (-0.25,0.7) {$\tilde{e}_{L}$};
			\node [right] at  (a3) {$u_{L}$};
			\node [right] at  (a4) {$e_{L}$};
			\node [right] at  (0.8,-0.4) {$e_{L}$};
			\node [right] at  (a7) {$u_{L}$};
			\node [left] at  (a2) {$d_{R}$};
			\node [left] at  (a6) {$d_{R}$};
			\node [above] at  (0.05,-0.7) {$\tilde{d}_{R}$};
			\node [left] at  (-0.25,0) {$\tilde{\chi}_{i}^{0}$};
			\node at (0,-1.35) {(c)};
		\end{tikzpicture}\\
		\begin{tikzpicture}[line width=1pt, scale=1.4,>=Stealth]
			\path (90:0.45) coordinate (a0);
			\path (90:1) coordinate (a1);
			\path (-0.8,1) coordinate (a2);
			\path (0.8,1) coordinate (a3);
			\path (0.8,0.45) coordinate (a4);
			\path (270:0.45) coordinate (a5);
			\path (-0.8,-1) coordinate (a6);
			\path (0.8,-1) coordinate (a7);
			\path (0.8,-0.45) coordinate (a8);
			\path (0.8,-1.4) coordinate (a9);
			\path (0,-1) coordinate (a10);
			\path (0,-1.4) coordinate (a11);
			\path (0,0.4) coordinate (a12);
			\path (0,0.0) coordinate (a13);
			\path (0,-0.4) coordinate (a14);
			\draw [scalarbar](a0)--(a1);
			\draw [fermion](a2)--(a1);
			\draw [fermion](a1)--(a3);
			\draw [fermion](a0)--(a4);
			\draw [fermion](a6)--(a10);
			\draw [fermion](a10)--(a7);
			\draw [scalarbar](a5)--(a10);
			\draw [fermion](a5)--(a8);
			\draw [fermionnoarrow](a13)--(a0);
			\draw [fermionnoarrow](a13)--(a5);
			\node [left] at  (90:0.7) {$\tilde{u}_{L}$};
			\node [right] at  (a3) {$e_{L}$};
			\node [right] at  (a4) {$u_{L}$};
			\node [right] at  (a8) {$u_{L}$};
			\node [right] at  (a7) {$e_{L}$};
			\node [left] at  (a2) {$d_{R}$};
			\node [left] at  (a6) {$d_{R}$};
			\node [left] at  (0,-0.7) {$\tilde{u}_{L}$};
			\node [left] at  (0,0) {$\tilde{\chi}_{i}^{0}$};
			\node at (0,-1.35) {(d)};
		\end{tikzpicture}\quad
		\begin{tikzpicture}[line width=1pt, scale=1.4,>=Stealth]
			\path (-0.25,0.45) coordinate (a0);
			\path (-0.25,1) coordinate (a1);
			\path (-0.8,1) coordinate (a2);
			\path (0.8,1) coordinate (a3);
			\path (0.8,0.45) coordinate (a4);
			\path (-0.25,-0.7) coordinate (a5);
			\path (-0.8,-0.7) coordinate (a6);
			\path (0.8,-1) coordinate (a7);
			\path (-0.25,0.0) coordinate (a13);
			\draw [scalarbar](a0)--(a1);
			\draw [fermion](a2)--(a1);
			\draw [fermion](a1)--(a3);
			\draw [fermion](a0)--(a4);
			\draw [fermion](-0.8,-0.7)--(-0.25,-0.7);
			\draw [scalarbar](0.25,-0.7)--(-0.25,-0.7);
			\draw [fermionbar](0.8,-1)--(0.25,-0.7);
			\draw [fermionbar](0.8,-0.4)--(0.25,-0.7);
			\draw [fermionnoarrow](a13)--(a0);
			\draw [fermionnoarrow](a13)--(a5);
			\node [left] at  (-0.25,0.7) {$\tilde{u}_{L}$};
			\node [right] at  (a3) {$e_{L}$};
			\node [right] at  (a4) {$u_{L}$};
			\node [right] at  (0.8,-0.4) {$e_{L}$};
			\node [right] at  (a7) {$u_{L}$};
			\node [left] at  (a2) {$d_{R}$};
			\node [left] at  (a6) {$d_{R}$};
			\node [above] at  (0.05,-0.7) {$\tilde{d}_{R}$};
			\node [left] at  (-0.25,0) {$\tilde{\chi}_{i}^{0}$};
			\node at (0,-1.35) {(e)};
		\end{tikzpicture}\quad
		\begin{tikzpicture}[line width=1pt, scale=1.4,>=Stealth]
			\path (-0.25,0.7) coordinate (a0);
			\path (-0.8,0.7) coordinate (a2);
			\path (0.8,1) coordinate (a3);
			\path (-0.25,-0.7) coordinate (a5);
			\path (-0.8,-0.7) coordinate (a6);
			\path (0.8,-1) coordinate (a7);
			\path (-0.25,0.0) coordinate (a13);
			\draw [scalarbar](0.25,0.7)--(-0.25,0.7);
			\draw [fermionbar](0.8,1)--(0.25,0.7);
			\draw [fermionbar](0.8,0.4)--(0.25,0.7);
			\draw [fermion](-0.8,0.7)--(-0.25,0.7);
			\draw [fermion](-0.8,-0.7)--(-0.25,-0.7);
			\draw [scalarbar](0.25,-0.7)--(-0.25,-0.7);
			\draw [fermionbar](0.8,-1)--(0.25,-0.7);
			\draw [fermionbar](0.8,-0.4)--(0.25,-0.7);
			\draw [fermionnoarrow](a13)--(a0);
			\draw [fermionnoarrow](a13)--(a5);
			\node [below] at  (0.05,0.7) {$\tilde{d}_{R}$};
			\node [right] at  (a3) {$u_{L}$};
			\node [right] at  (0.8,0.4) {$e_{L}$};
			\node [right] at  (0.8,-0.4) {$e_{L}$};
			\node [right] at  (a7) {$u_{L}$};
			\node [left] at  (a2) {$d_{R}$};
			\node [left] at  (a6) {$d_{R}$};
			\node [above] at  (0.05,-0.7) {$\tilde{d}_{R}$};
			\node [left] at  (-0.25,0) {$\tilde{\chi}_{i}^{0}$};
			\node at (0,-1.35) {(f)};
		\end{tikzpicture}\\
		\begin{tikzpicture}[line width=1pt, scale=1.4,>=Stealth]
			\path (90:0.45) coordinate (a0);
			\path (90:1) coordinate (a1);
			\path (-0.8,1) coordinate (a2);
			\path (0.8,1) coordinate (a3);
			\path (0.8,0.45) coordinate (a4);
			\path (270:0.45) coordinate (a5);
			\path (-0.8,-1) coordinate (a6);
			\path (0.8,-1) coordinate (a7);
			\path (0.8,-0.45) coordinate (a8);
			\path (0.8,-1.4) coordinate (a9);
			\path (0,-1) coordinate (a10);
			\path (0,-1.4) coordinate (a11);
			\path (0,0.4) coordinate (a12);
			\path (0,0.0) coordinate (a13);
			\path (0,-0.4) coordinate (a14);
			\draw [scalarbar](a0)--(a1);
			\draw [fermion](a2)--(a1);
			\draw [fermion](a1)--(a3);
			\draw [fermion](a0)--(a4);
			\draw [fermion](a6)--(a10);
			\draw [fermion](a10)--(a7);
			\draw [scalarbar](a5)--(a10);
			\draw [fermion](a5)--(a8);
			\draw [fermionnoarrow](a13)--(a0);
			\draw [fermionnoarrow](a13)--(a5);
			\node [left] at  (90:0.7) {$\tilde{u}_{L}$};
			\node [right] at  (a3) {$e_{L}$};
			\node [right] at  (a4) {$u_{L}$};
			\node [right] at  (a8) {$u_{L}$};
			\node [right] at  (a7) {$e_{L}$};
			\node [left] at  (a2) {$d_{R}$};
			\node [left] at  (a6) {$d_{R}$};
			\node [left] at  (0,-0.7) {$\tilde{u}_{L}$};
			\node [left] at  (0,0) {$\tilde{g}$};
			\node at (0,-1.35) {(g)};
		\end{tikzpicture}\quad
		\begin{tikzpicture}[line width=1pt, scale=1.4,>=Stealth]
			\path (-0.25,0.45) coordinate (a0);
			\path (-0.25,1) coordinate (a1);
			\path (-0.8,1) coordinate (a2);
			\path (0.8,1) coordinate (a3);
			\path (0.8,0.45) coordinate (a4);
			\path (-0.25,-0.7) coordinate (a5);
			\path (-0.8,-0.7) coordinate (a6);
			\path (0.8,-1) coordinate (a7);
			\path (-0.25,0.0) coordinate (a13);
			\draw [scalarbar](a0)--(a1);
			\draw [fermion](a2)--(a1);
			\draw [fermion](a1)--(a3);
			\draw [fermion](a0)--(a4);
			\draw [fermion](-0.8,-0.7)--(-0.25,-0.7);
			\draw [scalarbar](0.25,-0.7)--(-0.25,-0.7);
			\draw [fermionbar](0.8,-1)--(0.25,-0.7);
			\draw [fermionbar](0.8,-0.4)--(0.25,-0.7);
			\draw [fermionnoarrow](a13)--(a0);
			\draw [fermionnoarrow](a13)--(a5);
			\node [left] at  (-0.25,0.7) {$\tilde{u}_{L}$};
			\node [right] at  (a3) {$e_{L}$};
			\node [right] at  (a4) {$u_{L}$};
			\node [right] at  (0.8,-0.4) {$e_{L}$};
			\node [right] at  (a7) {$u_{L}$};
			\node [left] at  (a2) {$d_{R}$};
			\node [left] at  (a6) {$d_{R}$};
			\node [above] at  (0.05,-0.7) {$\tilde{d}_{R}$};
			\node [left] at  (-0.25,0) {$\tilde{g}$};
			\node at (0,-1.35) {(h)};
		\end{tikzpicture}\quad
		\begin{tikzpicture}[line width=1pt, scale=1.4,>=Stealth]
			\path (-0.25,0.7) coordinate (a0);
			\path (-0.8,0.7) coordinate (a2);
			\path (0.8,1) coordinate (a3);
			\path (-0.25,-0.7) coordinate (a5);
			\path (-0.8,-0.7) coordinate (a6);
			\path (0.8,-1) coordinate (a7);
			\path (-0.25,0.0) coordinate (a13);
			\draw [scalarbar](0.25,0.7)--(-0.25,0.7);
			\draw [fermionbar](0.8,1)--(0.25,0.7);
			\draw [fermionbar](0.8,0.4)--(0.25,0.7);
			\draw [fermion](-0.8,0.7)--(-0.25,0.7);
			\draw [fermion](-0.8,-0.7)--(-0.25,-0.7);
			\draw [scalarbar](0.25,-0.7)--(-0.25,-0.7);
			\draw [fermionbar](0.8,-1)--(0.25,-0.7);
			\draw [fermionbar](0.8,-0.4)--(0.25,-0.7);
			\draw [fermionnoarrow](a13)--(a0);
			\draw [fermionnoarrow](a13)--(a5);
			\node [below] at  (0.05,0.7) {$\tilde{d}_{R}$};
			\node [right] at  (a3) {$u_{L}$};
			\node [right] at  (0.8,0.4) {$e_{L}$};
			\node [right] at  (0.8,-0.4) {$e_{L}$};
			\node [right] at  (a7) {$u_{L}$};
			\node [left] at  (a2) {$d_{R}$};
			\node [left] at  (a6) {$d_{R}$};
			\node [above] at  (0.05,-0.7) {$\tilde{d}_{R}$};
			\node [left] at  (-0.25,0) {$\tilde{g}$};
			\node at (0,-1.35) {(i)};
		\end{tikzpicture}			
		\caption{The Feynman diagrams can contribute to $0\nu\beta\beta$ decay with the fermionic mediators to be neutralinos and gluino.}
		\label{SUSY-0vbb}
		  \end{figure}

\begin{figure}[t]
\centering
\includegraphics[height=4.8cm]{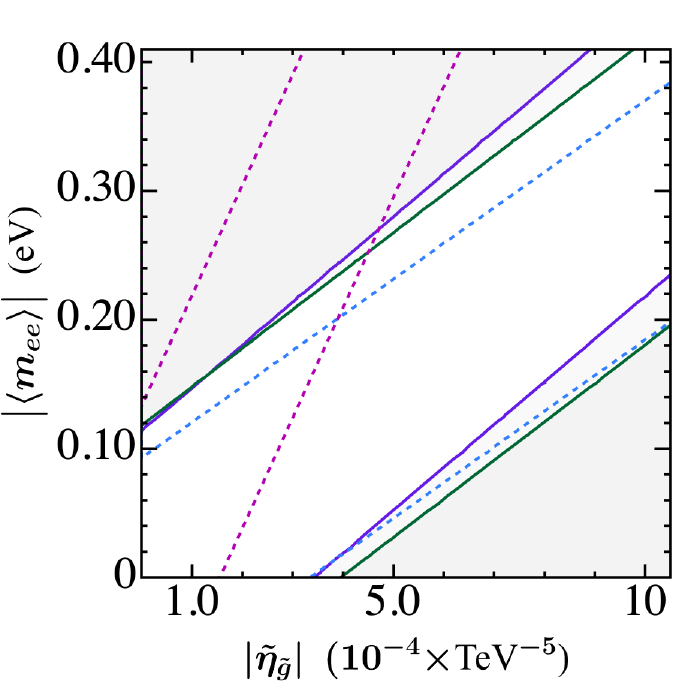}\quad
\includegraphics[height=4.8cm]{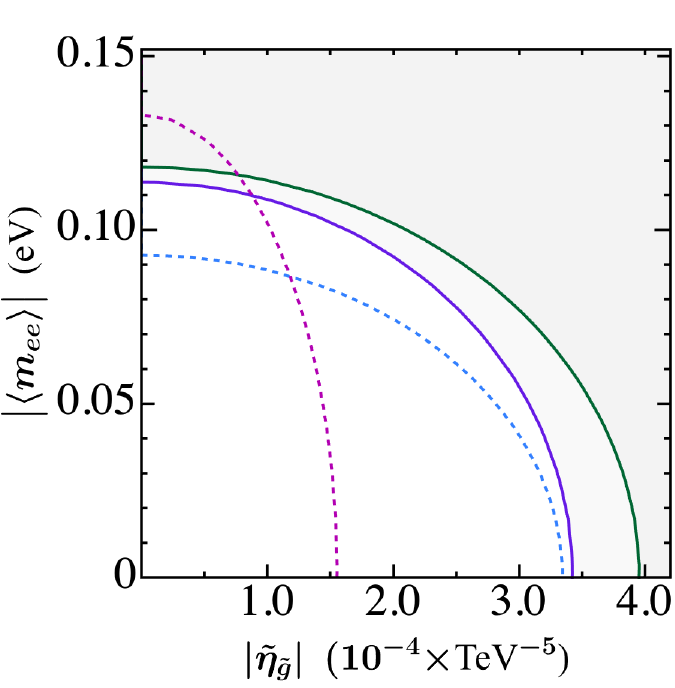}\quad
\includegraphics[height=4.8cm]{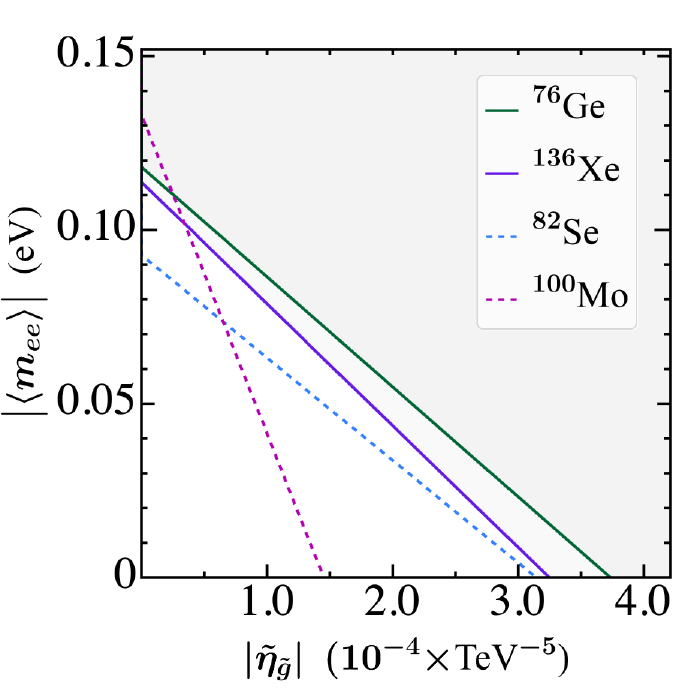}
\caption{The relation between the effective neutrino mass $\langle m_{ee}\rangle$ and the parameter $\tilde{\eta}_{\tilde{g}}=\lambda_{111}^{\prime 2}/m_{\tilde{g}}m_{\tilde{q}}^{4}$ is shown in these three figures, with the phase difference taken to be $0$ (left), $\pi/2$ (middle) and $\pi$ (right). The solid green and purple lines correspond to the GERDA and KamLAND-Zen experimental results, and the gray regions are excluded. The dashed blue and magenta lines correspond to the isotopes $^{82}$Se and $^{100}$Mo with the half-life set to be $10^{26}$ yrs and $3\times10^{26}$ yrs.}\label{SUSY-space}
\end{figure}

In the R-parity violating Supersymmetry ($\slashed{R}$-SUSY) model, the superpotential with R-parity violating that violates baryon number and lepton number contains the terms 
\begin{align}
W_{\slashed{R}}\supset\dfrac{1}{2}\lambda_{ijk}L_{i}L_{j}E_{k}^{c}+\lambda_{ijk}^{\prime}L_{i}Q_{j}D_{k}^{c}+\dfrac{1}{2}\lambda_{ijk}^{\prime\prime}U_{i}^{c}D_{j}^{c}D_{k}^{c}\,,
\end{align}
where $i,j,k$ are generation indices. The trilinear terms $\lambda_{ijk}$ and $\lambda_{ijk}^{\prime}$ lead to tiny neutrino mass at the one-loop level as shown in Fig.~\ref{SUSY-mv}.
The $\times$ denotes the insertion of Higgs bosons. The $\lambda_{111}^{\prime}$ terms also contribute to $0\nu\beta\beta$ decay where the fermionic mediator is neutralino or gluino~\cite{Mohapatra:1986su,Vergados:1986td,Hirsch:1995ek,Hirsch:1995cg}. The light neutralinos mediated case has been discussed in~\cite{Bolton:2021hje}. Here, the neutralinos, gluino, and sfermions are considered to be all heavy particles, with masses much larger than $\langle p^{2}\rangle\simeq$100 MeV that can induce the short-range mechanism. The Feynman diagrams are as shown in Fig.~\ref{SUSY-0vbb}.%
These Feynman diagrams can only contribute to the operators $\mathcal{O}_{1}^{RRL}$ and $\mathcal{O}_{2}^{RRL}$, so that one can write down the corresponding parameters $\epsilon_{1}^{RRL}$ and $\epsilon_{2}^{RRL}$ as
\begin{align}
\epsilon_{1}^{RRL}&=\dfrac{\eta_{a}}{8}-\dfrac{\eta_{b}}{16}-\dfrac{\eta_{c}}{16}+\dfrac{\eta_{d}}{32}+\dfrac{\eta_{e}}{32}+\dfrac{\eta_{f}}{32}+\dfrac{\eta_{g}}{24}+\dfrac{7\eta_{h}}{48}+\dfrac{\eta_{i}}{24}\,,\label{SUSY1}\\
\epsilon_{2}^{RRL}&=-\dfrac{\eta_{d}}{128}+\dfrac{\eta_{e}}{128}-\dfrac{\eta_{f}}{128}-\dfrac{\eta_{g}}{96}+\dfrac{\eta_{h}}{192}-\dfrac{\eta_{i}}{96}\label{SUSY2}\,,
\end{align}
and the $0\nu\beta\beta$ decay inverse half-life of the isotopes is given by
\begin{align}
T_{1/2}^{-1}=G_{11+}^{(0)}\left|\epsilon_{1}^{RRL}\mathcal{M}^{RRL}_{1}+\epsilon_{2}^{RRL}\mathcal{M}^{RRL}_{2}+\epsilon_{\nu}\mathcal{M}_{\nu}\right|^{2}\,.
\label{0vbb}
\end{align}
The corresponding dim-9 operators to the parameters in Eq.~(\ref{SUSY1}, \ref{SUSY2}) are
\begin{align}
&{\text{(a)}}\quad \dfrac{\eta_{a}}{8}\mathcal{O}_{1}^{RRL}\,,\quad \eta_{a}=4g^{2}\dfrac{\lambda_{111}^{\prime2}}{G_{F}^{2}V_{ud}^{2}}\sum\limits_{i}\dfrac{m_{p}}{m_{\tilde{\chi}_{i}^{0}}}\dfrac{(g_{ei}^{L})^{2}}{m_{\tilde{e}_{L}}^{4}}\,,\notag\\
&{\text{(b)}}\quad -\dfrac{\eta_{b}}{16}\mathcal{O}_{1}^{RRL}\,,\quad \eta_{b}=4g^{2}\dfrac{\lambda_{111}^{\prime2}}{G_{F}^{2}V_{ud}^{2}}\sum\limits_{i}\dfrac{m_{p}}{m_{\tilde{\chi}_{i}^{0}}}\dfrac{g_{ei}^{L}g_{ui}^{L}}{m_{\tilde{e}_{L}}^{2}m_{\tilde{u}_{L}}^{2}}\,,\notag\\
&{\text{(c)}}\quad -\dfrac{\eta_{c}}{16}\mathcal{O}_{1}^{RRL}\,,\quad \eta_{b}=4g^{2}\dfrac{\lambda_{111}^{\prime2}}{G_{F}^{2}V_{ud}^{2}}\sum\limits_{i}\dfrac{m_{p}}{m_{\tilde{\chi}_{i}^{0}}}\dfrac{g_{e\tilde{\chi}_{i}^{0}}^{LL}g_{d\tilde{\chi}_{i}^{0}}^{RR*}}{m_{\tilde{e}_{L}}^{2}m_{\tilde{d}_{R}}^{2}}\,,\notag\\
&{\text{(d)}}\quad \dfrac{\eta_{d}}{32}\mathcal{O}_{1}^{RRL}- \dfrac{\eta_{d}}{128}\mathcal{O}_{2}^{RRL}\,,\quad \eta_{d}=4g^{2}\dfrac{\lambda_{111}^{\prime2}}{G_{F}^{2}V_{ud}^{2}}\sum\limits_{i}\dfrac{m_{p}}{m_{\tilde{\chi}_{i}^{0}}}\dfrac{(g_{u\tilde{\chi}_{i}^{0}}^{LL})^{2}}{m_{\tilde{u}_{L}}^{4}}\,,\notag\\
&{\text{(e)}}\quad \dfrac{\eta_{e}}{32}\mathcal{O}_{1}^{RRL}+\dfrac{\eta_{e}}{128}\mathcal{O}_{2}^{RRL}\,,\quad \eta_{e}=4g^{2}\dfrac{\lambda_{111}^{\prime2}}{G_{F}^{2}V_{ud}^{2}}\sum\limits_{i}\dfrac{m_{p}}{m_{\tilde{\chi}_{i}^{0}}}\dfrac{g_{u\tilde{\chi}_{i}^{0}}^{LL}g_{d\tilde{\chi}_{i}^{0}}^{RR*}}{m_{\tilde{u}_{L}}^{2}m_{\tilde{d}_{R}}^{2}}\,,\notag\\
&{\text{(f)}}\quad\dfrac{\eta_{f}}{32}\mathcal{O}_{1}^{RRL}- \dfrac{\eta_{f}}{128}\mathcal{O}_{2}^{RRL}\,,\quad \eta_{f}=4g^{2}\dfrac{\lambda_{111}^{\prime2}}{G_{F}^{2}V_{ud}^{2}}\sum\limits_{i}\dfrac{m_{p}}{m_{\tilde{\chi}_{i}^{0}}}\dfrac{(g_{d\tilde{\chi}_{i}^{0}}^{RR*})^{2}}{m_{\tilde{d}_{R}}^{4}}\,,\notag\\
&{\text{(g)}}\quad\dfrac{\eta_{g}}{24}\mathcal{O}_{1}^{RRL}- \dfrac{\eta_{g}}{96}\mathcal{O}_{2}^{RRL}\,,\quad \eta_{g}=g_{s}^{2}\dfrac{\lambda_{111}^{\prime2}}{G_{F}^{2}V_{ud}^{2}}\dfrac{m_{p}}{m_{\tilde{g}}}\dfrac{1}{m_{\tilde{u}_{L}}^{4}}\,,\notag\\
&{\text{(h)}}\quad\dfrac{7\eta_{h}}{48}\mathcal{O}_{1}^{RRL}+ \dfrac{\eta_{h}}{192}\mathcal{O}_{2}^{RRL}\,,\quad \eta_{h}=g_{s}^{2}\dfrac{\lambda_{111}^{\prime2}}{G_{F}^{2}V_{ud}^{2}}\dfrac{m_{p}}{m_{\tilde{g}}}\dfrac{1}{m_{\tilde{u}_{L}}^{2}m_{\tilde{d}_{R}}^{2}}\,,\notag\\
&{\text{(i)}}\quad\dfrac{\eta_{i}}{24}\mathcal{O}_{1}^{RRL}- \dfrac{\eta_{i}}{96}\mathcal{O}_{2}^{RRL}\,,\quad \eta_{i}=g_{s}^{2}\dfrac{\lambda_{111}^{\prime2}}{G_{F}^{2}V_{ud}^{2}}\dfrac{m_{p}}{m_{\tilde{g}}}\dfrac{1}{m_{\tilde{d}_{R}}^{4}}\,,\notag
\end{align}
where $g,g_{s}$ are the coupling constants of $SU(2)_{L}$ and $SU(3)_{C}$, while $g_{\psi i}^{L(R)}$ is the coupling constant of left-handed (right-handed) fermion and sfermion with different neutralinos $\tilde{\chi}_{i}^{0}$~\cite{Haber:1984rc}. As~\cite{Hirsch:1995ek} has clarified if $m_{\tilde{q}}\sim m_{\tilde{e}},~m_{\tilde{\chi_{0}}}\geq 0.02 m_{\tilde{g}}$, the gluino contributions are much larger than the neutralinos contributions. The gluino contribution can be reorganized into
\begin{align}
  \eta_{g}=\eta_{h}=\eta_{i}=g_{s}^{2}\dfrac{\lambda_{111}^{\prime2}}{G_{F}^{2}V_{ud}^{2}}\dfrac{m_{p}}{m_{\tilde{g}}}\dfrac{1}{m_{\tilde{q}}^{4}}\equiv\tilde{\eta}_{\tilde{g}}\dfrac{g_{s}^{2}m_{p}}{G_{F}^{2}V_{ud}^{2}}\,,
\end{align}
 and the parameters are $  \epsilon_{1,\tilde{g}}^{RRL}=11\eta_{g}/48,
  \epsilon_{2,\tilde{g}}^{RRL}=-\eta_{g}/64$ under the assumption $m_{\tilde{u}_{L}}=m_{\tilde{d}_{R}}\equiv m_{\tilde{q}}$. The correlation between the $\langle m_{ee}\rangle$ and the parameter $\tilde{\eta}_{\tilde{g}}=\lambda_{111}^{\prime 2}/m_{\tilde{g}}m_{\tilde{q}}^{4}$ is shown in Fig.~\ref{SUSY-space}.~%
In the literature \cite{Agostini:2022bjh}, the authors have explored the potential of combining different experiments to distinguish the single gluino contribution. Here, with accurate NME calculations, we can conclude that future multipole isotope experiments could distinguish this model due to the distinct slope of the band in $^{100}$Mo compared to other isotopes.

\section{Discussion and Summary}\label{Sec5}
The $0\nu\beta\beta$ decay is the most promising way to probe the nature of neutrinos. This study set out to investigate the possibility of the attractive method that distinguishes the various Majorana neutrino mass models via the $0\nu\beta\beta$ decay using the isotopes $^{76}{\text{Ge}},^{82}{\text{Se}},^{100}{\text{Mo}}$ and $^{136}{\text{Xe}}$. The $0\nu\beta\beta$ decay can be described within the low-energy effective field theory framework, and the contributions come from the standard, long-range, and short-range mechanisms. The values of NMEs in the numerical analysis are taken within the IBM framework.

In a specific model, the $0\nu\beta\beta$ contributions usually come from multiple mechanisms. We analyze the scenario where the short-range and standard operators co-occur. It contains linear and elliptical cases based on the survival region's shape under the maximum interference assumption. The values of the slopes of the linear bands and tilt angles of the elliptics are listed in Table~\ref{linear-ratio1} and \ref{elliptical-list}. The corresponding contours of the effective neutrino mass and the Wilson coefficients are shown in Fig.~\ref{linear-plot} and \ref{elliptical-plot}.

Specific Majorana neutrino mass models are revisited to figure out whether the complementary searches in multiple isotopes can distinguish the models or examine the parameter survival regions. The discussion includes the tree-level Type-I seesaw model, Type-II seesaw model, and left-right symmetric model, which can also be divided into Type-I dominance and Type-II dominance scenarios. Two models that generate tiny neutrino mass via one loop are focused, one contains the leptoquarks, and the other is the $\slashed{R}$-SUSY model. The correlations between effective neutrino mass and parameters with the interference effects considered are also shown.

For future $0\nu\beta\beta$ decay experiments in different isotopes, the following scenarios may arise with the assumption that the sensitivities of the experiments are at comparable levels:
\begin{itemize}
\item[$\bullet$]~No signal is detected in experiments with any isotopes.

The experimental limits indicate that the outer areas of the lines in the figures can be ruled out. The inner area of the line, also called the survival region, with a specific isotope can be the shape of a band or an elliptic. Within two different isotopes, the survival region can be reduced to the overlap area. The greater the discrepancy between those values of the slopes or the tilt angles in different isotopes, the narrower the overlap area becomes. Considering more than two isotopes, the overlap area can be further reduced. A notable example is the case corresponding to $\mathcal{O}_{2}$ in Fig.~\ref{linear-plot}. The combination of experiment searches allows for a more comprehensive examination of the contours of the effective neutrino mass and parameters. This conclusion is universal for all cases in operator analysis and all models, except for the Type-II seesaw model. 

\item[$\bullet$]~Signals are observed in experiments with a specific isotope, but no signal is found in experiments with other isotopes.

In most of the cases in operator analysis and the models, the differences between the slopes or the tilt angles in $^{76}\text{Ge}$, $^{82}\text{Se}$ and $^{136}\text{Xe}$ isotopes are small. When the sensitivities of the experiments are close, this scenario may predict that some cases can be excluded.

Taking Type-I seesaw as an example, the slopes or tilt angles have similar values in $^{76}\text{Ge}$, $^{82}\text{Se}$ and $^{136}\text{Xe}$ isotopes, no matter how great the interference effects are, as shown in Fig.~\ref{seesawType-I}. This suggests that if a signal can be detected in the $^{82}\text{Se}$ experiments, similar signals should also be observed in the $^{76}\text{Ge}$ and $^{136}\text{Xe}$ experiments at comparable levels of sensitivities. Therefore, if one of these isotopes has experimental signals while others do not, there is a high probability that this model cannot be a valid explanation. 

In the cases corresponding to $\mathcal{O}_{1,5}$ and the $\slashed{R}$-SUSY model, there are distinctive differences between the slopes or the tilt angles in $^{100}\text{Mo}$ and the other three isotopes. The signal in the future $^{100}\text{Mo}$ experiments lies along the boundary of the band and the elliptic. With the null decay signals in other isotopes, the boundary can be examined more comprehensively. Using the $\slashed{R}$-SUSY model as an instance, if there is a signal with the half-life to be $3\times10^{26}$ yrs in $^{100}\text{Mo}$ which refers to the dashed magenta curves in Fig.~\ref{SUSY-space}, part of the curves can be excluded by the current GERDA and KamLAND-Zen experiments. 

Moreover, a bold corollary is that if a signal is observed in $^{100}\text{Mo}$ experiment, while no signal in other isotopes with similar sensitivity, it will be more likely to reveal underlying contributions from $\mathcal{O}_{1,5}$ or from the $\slashed{R}$-SUSY model.

\item[$\bullet$]~Signals are observed in experiments within more than one isotope.

As we have discussed before, the signal in one isotope corresponds to a curve in the correlation figures. If there are signals in experiments within two different isotopes, the intersections of the two curves can be the solutions. The values of the effective neutrino mass and the Wilson coefficients or parameters in models can be determined. The null results in other isotopes can help to examine the solutions. 

Furthermore, if the decay signals are observed in more than two isotopes, things become more interesting. For instance, in the case corresponding to $\mathcal{O}_{2}^{XXL}$ as shown in Fig.~\ref{linear-plot}, the slopes are distinctive in different isotopes. Finding signals in three or more isotopes results in their intersections being very close to where the Wilson coefficient is zero. This indicates that this case can be highly likely excluded.

As the half-life of an isotope $0\nu\beta\beta$ decay has a fixed value, once the signal is observed, the corresponding curve will be determined. If the curves in different isotopes have no intersections, the cases or the models can be ruled out. This scenario is more likely to occur in those cases and models where the slopes and the tilt angles are similar in different isotopes. 
\end{itemize}

In conclusion, whether the slopes and the tilt angles have distinctive values in different isotopes can be used as a good criterion to determine the abilities of experiments to distinguish operators and models. Different scenarios of future experiments within multiple isotopes can reveal different contributions from operators and models. Future experiments based on different isotopes will be beneficial to have a deeper understanding of the $0\nu\beta\beta$ decay mechanisms and the underlying new physics contributions.

\acknowledgments
We are grateful to Jiang-Ming Yao and Chen-Rong Ding for useful discussions. This work is supported in part by the National Science Foundation of China (12175082) and the Fundamental Research Funds for the Central Universities (2023CXZZ132). 

\bibliographystyle{JHEP}
\bibliography{references}

\end{document}